\documentclass[11pt,psfig,a4]{amsart}

\usepackage{nccmath}
\usepackage{mathrsfs}
\usepackage{amssymb}
\usepackage{geometry}
\usepackage{graphics}
\usepackage[pdftex]{hyperref}

\usepackage[dvips]{graphicx}
\usepackage{pgf}
\usepackage{tikz}
\usepackage{amsthm}

\usetikzlibrary{arrows,automata,positioning}

\tikzstyle{vertex}=[circle, draw, inner sep=0pt, minimum size=4pt]

\tikzset{main node/.style={circle,fill=blue!20,draw,minimum size=1cm,inner sep=0pt},
            }

\newtheorem{maintheorem}{Theorem}[section]
\newtheorem{theorem}{Theorem}[section]

\newtheorem{pro}[theorem]{Proposition}
\newtheorem{cor}[theorem]{Corollary}

\theoremstyle{definition}

\newtheorem{example}[theorem]{Example}

\theoremstyle{remark}

\numberwithin{equation}{section}

\begin{document}
\pagestyle{plain}

\def\beq{\begin{equation}}
\def\eeq{\end{equation}}
\def\eps{\epsilon}
\def\laa{\langle}
\def\raa{\rangle}
\def\qed{\begin{flushright} $\square$ \end{flushright}}
\def\qee{\begin{flushright} $\Diamond$ \end{flushright}}
\def\ov{\overline}
\def\bma{\begin{bmatrix}}
\def\ema{\end{bmatrix}}

\def\ora{\overrightarrow}

\def\bma{\begin{bmatrix}}
\def\ema{\end{bmatrix}}
\def\bex{\begin{example}}
\def\eex{\end{example}}
\def\beq{\begin{equation}}
\def\eeq{\end{equation}}
\def\eps{\epsilon}
\def\laa{\langle}
\def\raa{\rangle}
\def\qed{\begin{flushright} $\square$ \end{flushright}}
\def\qee{\begin{flushright} $\Diamond$ \end{flushright}}
\def\ov{\overline}

\author{Carlos F. Lardizabal}

\address{Instituto de Matem\'atica e Estat\'istica, Universidade Federal do Rio Grande do Sul - UFRGS. Av. Bento Gon\c calves 9500 - CEP 91509-900, Porto Alegre, RS, Brazil}
\email{cfelipe@mat.ufrgs.br}

\title{Mean hitting times of quantum Markov chains in terms of generalized inverses}

\begin{abstract}
We study quantum Markov chains on graphs, described by completely positive maps, following the model due to S. Gudder (J. Math. Phys. 49, 072105, 2008) and which includes the dynamics given by open quantum random walks as defined by S. Attal et al. (J. Stat. Phys. 147:832-852, 2012). After reviewing such structures we examine a quantum notion of mean time of first visit to a chosen vertex. However, instead of making direct use of the definition as it is usually done, we focus on expressions for such quantity in terms of generalized inverses associated with the walk and most particularly the so-called fundamental matrix. Such objects are in close analogy with the theory of Markov chains and the methods described here allow us to calculate examples that illustrate similarities and differences between the quantum and classical settings.
\end{abstract}

\maketitle

\date{\today}

\section{Introduction}

In quantum information theory one is often interested in the problem of finding quantum versions of classical probabilistic notions. Among the several well-known statistical objects which can be studied, one may consider {\it Markov chains}  \cite{aldous,bremaud,snell,peres}. The importance of such random processes is due to their elegant mathematical structure and the great variety of concrete phenomena which can be modeled by them. 

\medskip

We note that the term {\it quantum Markov chain  (QMC)} has been employed in more than one context motivated by quantum mechanics \cite{accardi2, accardi1,dhahri,gudder,gudder2,sutter}. In this work we will focus on the model described by S. Gudder \cite{gudder}: consider a graph on which a quantum particle is located, with such particle possessing some internal degree of freedom (e.g. spin). A transition from some chosen vertex $j$ to another vertex $i$ is described by an effect operator given by a completely positive (CP) map acting on positive semidefinite matrices \cite{nielsen}. The map formed by the collection of all such effect operators is then called a QMC. 

\medskip

We note that the QMC model mentioned above includes the class of {\it open quantum random walks (OQWs)} \cite{attal} and, as the name suggests, such walks are not described by unitary operators in general. Instead, they are meant to describe dissipative quantum dynamics \cite{breuerp}. On one hand, due to their specific properties, unitary quantum walks have attracted much attention in the study of quantum computing, most particularly in search algorithms \cite{portugal}, simulations of complex systems \cite{kassal}, the study of topological phases \cite{kitagawa} and other topics \cite{salvador}. On the other side, dissipative quantum walks on graphs seem to have received much less attention, so it is one of the goals of this work to examine a set of questions in such context. The problems of interest in this article will concern the calculation of mean time of first visit to a vertex, a well-studied topic in the unitary context \cite{ambainis2}, but not so much in open quantum settings.

\medskip

From a theoretical point of view, QMCs stand in an intermediate position between a) the classical theory of Markov chains/stochastic processes associated with contractive operators and b) the theory of closed quantum dynamics associated with unitary operators. As a consequence, the statistics obtained for QMCs will resemble some kind of classical information which has been modified by the knowledge of the internal degree of freedom of the particles involved.

\medskip

It is worth noting that problems in open quantum systems are of great relevance in the study of quantum information and quantum computation: the physical and mathematical properties of noisy communication channels and the potential usefulness of decoherence appearing in certain quantum systems are seen to be important topics of discussion \cite{verst}. In the context of OQWs, see \cite{sinayskiy0} regarding  implementations of dissipative quantum computing algorithms (these being illustrated by the Toffoli gate and the Quantum Fourier Transform) and we refer the reader to \cite{spsurvey} for a survey of recent results on the field. In this work we will study both QMCs and OQWs, but giving emphasis on the former (both notions are carefully defined in this work). We will be interested in understanding which parts of the theoretical formalism can be obtained from well-known classical structures and how much needs to be modified in order to describe phenomena which are typical of quantum information processing.

\section{Motivation and statement of the problems: hitting times and generalized inverses}

The problem of estimating the mean times of first visit and return in quantum settings has been addressed in recent years, this being in part motivated by the study of quantum walks, their application to search algorithms and related topics on quantum information processing \cite{accardi1,kempe2,brun0,portugal}. One notices that there is more than one quantum notion for the time of first visit, each with its own features, so establishing a consistent mathematical framework where problems of interest can be posed properly is an important first step. 

\medskip

In this work we are interested in further studying the problem of mean hitting time to reach some chosen vertex on a graph for a particle evolving under a QMC dynamics. Mean hitting times for OQWs have been studied in \cite{ls2016,oqwmhtf,gawron} and we note that although our focus is on QMCs, the model of OQWs will serve as a good guide for the general case discussed here.

\medskip

We are in part motivated by the classical problem of calculating mean hitting times for a walker on a graph under a Markov chain dynamics: given a graph and transition probabilities between its vertices, what is the mean time for a walker to reach vertex $j$ for the first time, given that it has started at vertex $i$? Formally, the (classical) mean hitting time is given by
$$E_i(T_j)=\sum_t tP_i(T_j=t)$$
where $T_j$ is the random variable given by the time of first visit to vertex $j$, and $P_i(T_j=t)$ is the probability that $T_j=t$, given that the walk begins at position $i$. 

\medskip

From a theoretical point of view there is interest in obtaining methods for calculating hitting times which avoid making direct use of the definition \cite{snell,peres,wang}. From the theory of Markov chains we know that, alternatively, the mean hitting time can be calculated via the so-called {\it fundamental matrix} associated with a finite ergodic Markov chain with stochastic matrix $P$,
$$Z=(I-P+\Omega)^{-1}$$
where $\Omega=\lim_{m\to\infty} P^m$, and for which the following equation is valid:
\beq\label{clmhtf} E_i(T_j)=\frac{Z_{jj}-Z_{ij}}{\pi_j}\eeq
Above $\pi=(\pi_i)$ denotes the unique fixed probability associated with the walk. This is the {\it mean hitting time formula (MHTF)}\cite{aldous,bremaud}, and is one of several expressions relating $Z$ with statistical quantities of the associated walk. 

\medskip

Regarding the fundamental matrix we draw attention to an important aspect of such map, namely, that $Z$ is a particular example of a {\it generalized inverse} of the operator $I-P$ \cite{hunter,meyer75}. As it is well-known, such inverses are in general nonunique and possess a rich algebraic theory in connection with applications to stochastic processes, linear estimation, difference equations  and other areas \cite{wang}. In the setting of Markov chains, it is known that the group (Drazin) inverse enjoys a central role: if $P$ denotes a finite stochastic matrix, $A=I-P$ and if $A^\#$ denotes its group inverse, then
$$\lim_{m\to\infty}\frac{I+P+P^2+\cdots+P^{m-1}}{m}=I-AA^\#$$
and several related results hold if one assumes, for instance, regularity or that the chain is absorbing. As one may guess, the fact that there are several kinds of generalized inverses suggests that some of them may be more or less suitable to a given problem. This is the case of the group inverse with respect to Markov chains, since it is the one which leads to greater practical and computational advantage \cite{campbell}. With respect to the problem of calculating mean hitting times, Hunter \cite{hunter} has shown the usefulness of such inverses in the context of discrete and continuous time Markov chains.

\medskip

The previous discussion leads to the following natural questions:

\begin{enumerate}
\item Regarding a setting for which one has a quantum particle on a graph and by considering a quantum channel which dictates the statistical behavior of its position, is there an associated hitting time formula?
\item In the context of QMCs, is there an algebraic structure coming from generalized inverses which is closely related to mean hitting times? 
\end{enumerate}

Questions 1 and 2 above will be answered in the affirmative and formally proven in later sections. The importance of such problems is in part justified by the following remarks:

\medskip

a) Generalized inverses play a comprehensive role in the study of classical Markov chains: given a random walk, essentially every quantity of interest can be described in terms of such inverses \cite{meyer75}, with mean hitting times being one of the most representative examples. Having in mind this nontrivial mathematical structure for classical processes, it is natural to ask what is available in the quantum setting. The practical use of generalized inverses in quantum walk settings seems to be an unexplored topic and the existence of hitting time formulae (i.e., a formula providing mean hitting times in terms of a generalized inverse) seems to be an almost untouched problem so far. It is our perception that the search of such expressions quantum contexts may provide an improved understanding of relevant aspects of QMCs and quantum walks in general.

\medskip

b) The existence of hitting time formulae  in quantum contexts may offer a new understanding  regarding the quadratic gain in time that certain quantum search algorithms present over classical counterparts. For instance, regarding the Grover algorithm \cite{nielsen,portugal}, while it is true that such procedure occurs in an unitary setting, one can also discuss search algorithms executed by dissipative quantum channels and study their performance \cite{ellinas}. The potential usefulness of noisy channels for quantum information processing has been discussed in the literature \cite{kendon,verst} so it is a problem of interest to understand, in formal terms, the behavior of QMCs in terms of decoherences applied to some unitary dynamics. 

\medskip

In a similar vein, regarding walks on the integer lattices, we recall that the expected distance from the origin is measured by the standard deviation of the probability distribution. The quantum walk on both the line and the two-dimensional lattice has a standard deviation that is directly proportional to the evolution time, this being in contrast to the standard deviation of the classical random walk, which is proportional to the square root of the evolution time \cite{portugal}. Then it is a relevant question to ask whether some hitting time formula sheds some light on this matter, perhaps in terms of some aspect of the generalized inverse appearing in such expressions.

\medskip

For potential applications of hitting times in the unitary context we refer the reader to \cite{ambainis2,kempe2}, regarding quantum search algorithms, and in the context of OQWs we refer the reader to \cite{oqwmhtf} and references therein, noting that to a great extent QMCs are also contemplated on such matters.

\section{Overview of results and structure of the work}

In this work we establish the following results:

\medskip

1. {\it Mean hitting time formula (MHTF) for ergodic QMCs in terms of generalized inverses.} One of the main objectives of this work is to illustrate how generalized inverses can be used to study mean hitting times of QMCs. In particular we will show how a classical proof can be modified in order to produce a result in an open quantum setting. Our main result describes a method of calculating the mean time of first visit to a vertex in terms of arbitrary generalized inverses, a result we call {\it Hunter's formula for QMCs} (Theorem \ref{bigtheorem}). We review the theory of generalized inverses, noting that in this part we are closely motivated by the classical reference due to Hunter \cite{hunter}. The QMC is assumed to be ergodic, which corresponds to the assumption of aperiodicity and irreducibility, these resembling the usual Markov chain hypothesis for the corresponding classical result (these are reviewed later in this work).

\medskip

2. {\it Mean hitting time formulae in terms of the fundamental matrix.} A hitting time formula has been proved in \cite{oqwmhtf} for finite irreducible unital OQWs. This will be reviewed later noting that such result can be generalized, namely, the formula holds for finite QMCs, including the nonunital case. This is Theorem \ref{oqwmhtfstat}, which we call the {\it first MHTF}. Theorem \ref{MHTF2} is called the {\it second MHTF} and concerns the mean hitting time for reaching a vertex for the first time, given some initial vertex randomly chosen according to the stationary density of the channel. Both results concern the case for which the generalized inverse chosen is the fundamental matrix. We believe some of the techniques employed here will be seen to be of independent interest to the quantum information community.  The results in this part are motivated by \cite{oqwmhtf} and \cite{peres}.

\medskip

3. {\it Computing concrete examples.} In Section \ref{secex} we illustrate the theory with examples. At this point we are able to take a brief look at the large set of possibilities for choosing a generalized inverse associated to a QMC. In particular,  Example \ref{exemplo2} is such that the entries of the QMC depend on a scalar parameter $p$ and it is explained that the entries of certain generalized inverses can be written as rational functions of $p$ (we conjecture that this is the case for every generalized inverse). However, the degree of the polynomial numerators and denominators seems to depend on the particular choice of inverse. This and other algebraic aspects may lead to other questions, not only with respect to analogies with the classical case, but also regarding the important and nontrivial case of unitary dynamics. 

\medskip

Finally, it is worth to point out the following: we do not claim that hitting times are somehow easier to calculate via generalized inverses than by using the definition directly. Instead, our focus is on the existence of the inverses and their applications. Although the practical calculation of large QMC examples may be difficult in general, several auxiliary results can be performed with the use of a computer and may lead to useful information regarding comparisons with their classical counterparts.

\medskip

The structure of this work is as follows: in Section \ref{sec2} we review the QMC and OQW models and define the probability notions which will be studied. In Section \ref{sec3} we define the block  matrix representation on which both theory and examples will rely on, highlighting similarities and differences with respect to the classical case of stochastic matrices.  Section \ref{sec4} and \ref{sec5} present the hitting time formulae according to the above statements and Section \ref{secex} describes the examples. In Section \ref{sec6} we conclude by briefly discussing possible research directions. The Appendix presents the proofs of the results. 

\section{Preliminaries}\label{sec2}

\subsection{Quantum Markov chains on a finite graph}

Consider the set
$$\mathcal{D}_{n;k}:=\{\rho=[\rho_1\;\cdots\;\rho_n]^T:\; \rho_i\in M_k(\mathbb{C}), \;\rho_i\geq 0, \;i=1,\dots,n,\; \sum_{j=1}^n \mathrm{Tr}(\rho_j)=1\}$$
where $M_k(\mathbb{C})$ denotes the order k complex matrices and $\rho_i\geq 0$ means that $\rho_i$ is positive semidefinite. We call $n$ the number of vertices and $k$ the internal degree of freedom. Now let
$$\Phi_i(\rho):=\sum_{j=1}^n \Phi_{ij}(\rho_j),\;\;\;\rho\in\mathcal{D}_{n;k},\;\;\;i=1,\dots,n$$
where each $\Phi_{ij}$ is a CP map, and
\beq\label{defqmcgen} T(\rho)=\begin{bmatrix} {\Phi}_{11} & \cdots & {\Phi}_{1n} \\ {\Phi}_{21} & \cdots &  {\Phi}_{2n} \\ \vdots & \ddots & \vdots \\ {\Phi}_{n1} & \cdots & {\Phi}_{nn}\end{bmatrix}\cdot\begin{bmatrix} \rho_1 \\ \rho_2 \\ \vdots \\ \rho_n\end{bmatrix}:=\begin{bmatrix} \Phi_1(\rho)\\\Phi_2(\rho) \\ \vdots \\ \Phi_n(\rho)\end{bmatrix},\;\;\;\rho\in\mathcal{D}_{n;k}\eeq
and we assume trace preservation, that is,
$$\mathrm{Tr}\Big(\sum_{i=1}^n \Phi_{i}(\rho)\Big)=\mathrm{Tr}(\rho),\;\;\;j=1,\dots,n,\;\;\;\rho\in\mathcal{D}_{n;k}$$
We will call the CP map defined by the above operator matrix $T$ a {\it quantum Markov chain (QMC)} \cite{gudder}, see Figure \ref{fig0}. We say a QMC is {\it finite} if it acts on a finite graph, as it will be the case in this entire work. For concrete examples of graphs and transition effects, we refer the reader to the examples in Section \ref{secex}. A density $\rho\in\mathcal{D}_{n;k}$ will be called a {\it QMC density}.

\medskip

\begin{center}
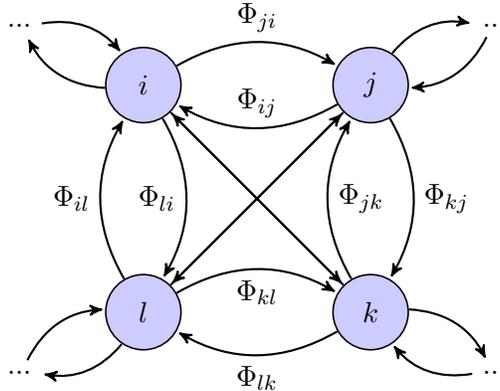
\begin{figure}[ht]
\begin{tikzpicture}
[->,>=stealth',shorten >=1pt,auto,node distance=2.0cm,
                    semithick]
    \node[main node] (1) {$i$};
    \node[main node] (2) [right=2.0cm and 2.0cm of 1,label={[]$$}]  {$j$};

    \node[main node] (3) [below=2.0cm and 2.0cm of 2,label={[]$$}]  {$k$};
    \node[main node] (4) [left = 2.0cm and 2.0cm of 3] {$l$};

    \node[] (1a) [left=1.0cm and  1.0cm of 1] {};
    \node[] (1b) [above=0.5cm and  0.5cm of 1a] {$...$};

    \node[] (2a) [right=1.0cm and  1.0cm of 2] {};
    \node[] (2b) [above=0.5cm and  0.5cm of 2a] {$...$};

    \node[] (3a) [right=1.0cm and  1.0cm of 3] {};
    \node[] (3b) [below=0.5cm and  0.5cm of 3a] {$...$};

    \node[] (4a) [left=1.0cm and  1.0cm of 4] {};
    \node[] (4b) [below=0.5cm and  0.5cm of 4a] {$...$};



    \path[draw,thick]


    (1) edge   [bend left]  node {$\Phi_{ji}$}    (2)
    (1) edge   [bend left]  node[above]{$$}   (1b)
    (1b) edge   [bend left]  node[below] {$$}    (1)

    (2) edge   [bend left]     node[above]{$\Phi_{ij}$} (1)
    (2) edge   [bend left]  node[above]{$$}   (2b)
    (2b) edge   [bend left]  node[below] {$$}    (2)

    (3) edge   [bend left]  node[above]{$$}   (3b)
    (3b) edge   [bend left]  node[below] {$$}    (3)

    (4) edge   [bend left]  node[above]{$$}   (4b)
    (4b) edge   [bend left]  node[below] {$$}    (4)

    (2) edge   [bend left]     node {$\Phi_{kj}$} (3)
    (3) edge   [bend left]     node[right] {$\Phi_{jk}$} (2)

    (1) edge   [bend left]     node[left] {$\Phi_{li}$} (4)
    (4) edge   [bend left]     node[left] {$\Phi_{il}$} (1)

    (4) edge   [bend left]     node[below] {$\Phi_{kl}$} (3)
    (3) edge   [bend left]     node[below] {$\Phi_{lk}$} (4)


    (3) edge      node {$$} (1)
    (1) edge        node {$$} (3)

    (2) edge      node {$$} (4)
    (4) edge        node {$$} (2)



    ;
\end{tikzpicture}
\caption{Schematic illustration of quantum Markov chains. The walk is realized on a graph with a set of vertices denoted by $i,j,k,l,\dots$ and each operator $\Phi_{ij}$ is a completely positive map describing a transformation in the internal degree of freedom of the particle during the transition from vertex $j$ to vertex $i$. For simplicity of illustration some edges are not labeled. In the particular case that all maps are conjugations, i.e., for every $i,j$, $\Phi_{ij}=B_{ij}\cdot B_{ij}^\dagger$ for certain matrices $B_{ij}$ the quantum Markov chain is called an open quantum random walk.}\label{fig0}
\end{figure}
\end{center}

\medskip

An important special case of QMC is given by the following. Assume that above we have
$$\Phi_{ij}(\rho)=B_{ij}\rho_j B_{ij}^\dagger,\;\;\;B_{ij}\in M_k(\mathbb{C}),\;\;\;\rho\in\mathcal{D}_{n;k},\;\;\;i=1,\dots,n$$
We call $B_{ij}$ the effect matrix of transition from vertex $j$ to vertex $i$. Once again we have a CP map: if $\{|i\rangle: i=1,\dots,n\}$ is an orthonormal basis for $\mathbb{C}^n$, the above is equivalent to write
$$T(\rho)=\sum_{i,j=1}^n M_{ij}\rho M_{ij}^\dagger,\;\;\;M_{ij}=B_{ij}\otimes|i\rangle\langle j|$$
with the densities of $\mathcal{D}_{n;k}$ being written as $\rho=\sum_{i=1}^n \rho_i\otimes|i\rangle\langle i|$. 
We will be interested in the trace-preserving case, that is, we will assume that
$$\sum_{i=1}^n B_{ij}^\dagger B_{ij}=I,\;\;\;j=1,\dots,n$$
Under such condition it is clear that $\mathcal{D}_{n;k}$ is preserved by $T$ and in this case such map will be called an {\it open quantum random walk} on a graph with $n$ vertices, internal degree of freedom $k$ and transition effect matrices $B_{ij}$, $i,j=1\dots,n$ \cite{attal}. We emphasize that in this work we will only consider the statistics of iterative quantum trajectories \cite{kummerer}, meaning that we are summing square moduli of amplitudes (traces) associated with individual paths. This is in contrast to the case of unitary quantum random walks \cite{werner,portugal}, for which one calculates probabilities via the square modulus of a sum of amplitudes. We also note that the case $k=1$ reduces to the classical case, that is, $T$ becomes an order $n$ column stochastic matrix acting on the set of probability vectors on $\mathbb{C}^n$. 

\medskip

Other topics on OQWs which have been examined are the following: reducibility, periodicity, ergodic properties \cite{cp}; large deviations \cite{cp2}; open quantum Brownian motions \cite{sinayskiy4}; site (vertex) recurrence of OQWs \cite{bbp,cgl,dhahri,ls2015}. We refer the reader to \cite{spsurvey} for a recent survey. 

\medskip

We also recall that if $(\rho_n)_{n\in\mathbb{N}}$ is the sequence of densities obtained by repeated measurements of a OQW and $(X_n)_{n\in\mathbb{N}}$ denotes the position of such densities at time $n$ then the process $(\rho_n,X_n)_{n\in\mathbb{N}}$ is in fact a Markov chain in the classical sense (see \cite{attal,attal2,cp2} for more details). However, we are often interested in facts which do not follow immediately from such property and which do not have a classical correspondent. For instance, the sequence of positions $(X_n)_{n\in\mathbb{N}}$ alone is {\it not} a Markov chain, since one also needs to carry the information on the internal degree of freedom at all times in order to calculate probabilities in a Markovian way.  This has implications, for instance, on the problem of site recurrence for OQWs (see \cite{bbp,cgl}) and also on the problem of first visit to a vertex \cite{ls2016,oqwmhtf}, which is studied in this work. 

\medskip

Finally, it is worth noting the basic distinction between QMCs and the well-known {\it coined unitary quantum walks} \cite{ambainis,kempe,portugal,salvador}. We recall that the unitary walk acting on the integers can be written as
$$U=S\cdot(C\otimes I)$$
where $U$ is a linear unitary operator acting on the Hilbert space given by the tensor product of the so-called coin space and the state space. The unitary map $C$ is defined as the coin, and the map $S$ is the shift operator. For 1-qubit walks we can write
$$C=\begin{bmatrix} a & b \\ c & d \end{bmatrix}, \;\;\; R=\begin{bmatrix} a & b \\0 & 0 \end{bmatrix},\;\;\; L=\begin{bmatrix} 0 & 0 \\ c & d \end{bmatrix}$$
and assume that at time $t=0$ we prepare a state $|\psi\rangle$ localized at $|0\rangle$. Then the probability of finding the particle at a certain position $X_n=|i\rangle$ at time $n$ is given by a {\it square modulus of a sum of amplitudes}, that is,
$$\mathbb{P}_{|\psi\rangle\otimes|0\rangle}(X_n=|i\rangle)=\Bigg\Vert\sum_{\{i_1,\dots,i_n\}\in \mathcal{P}(|0\rangle\to |i\rangle)} C_{i_1}\cdots C_{i_n}|\psi\rangle\Bigg\Vert^2$$
where $\mathcal{P}(|0\rangle\to |i\rangle)$ is the collection of all sequences of numbers $i_k\in\{\pm 1\}$ such that $\sum_k i_k=i$ and write $C_1=L$, $C_{-1}=R$ from which we obtain products of matrices associated with paths beginning at $|0\rangle$ and ending at $|i\rangle$. The resulting statistics are given by the Konno distribution \cite{konno}. On the other hand, for a QMC procedure we prepare an initial density matrix, apply the channel on it and then perform a measurement to determine the vertex for which the system has evolved to. Then by normalizing we obtain a new density and we repeat the process \cite{attal,cp}. So, at each step we have a probability computation and this procedure is thus identified with {\it a summation of square modulus of amplitudes}. In this case each modulus can be written as a trace calculation and as a result we obtain Gaussian curves or linear combinations of these as probability distributions. More precisely, it has been proved that a Central Limit Theorem holds for OQWs \cite{attal2,sadowski}, a fact which is in clear contrast with the behavior of unitary walks \cite{ambainis,portugal}.

\subsection{Probability notions}

We note that when we refer to the position of a quantum particle on a graph, it is implicit that we are considering a {\it monitoring procedure} \cite{bourg,gv,werner}, that is, we inspect whether the particle has been found at some chosen vertex. If the answer is positive, the experiment is over. Otherwise the particle is in the subspace given by the complement associated with the vertex inspected, the walk continues and we repeat the process iteratively. With respect to such formalism, we define the following probabilistic quantities for a QMC:
\begin{flalign}
& p_r(\rho\to j)=\text{probability of reaching vertex $j$ in $r$ steps when starting at the state $\rho$.} \nonumber\\
& \pi_r(\rho\to j)=\text{probability of reaching vertex $j$ for the first time in $r$ steps when starting at the state $\rho$.} \nonumber\\
& \pi(\rho\to j)=\text{probability of ever reaching vertex $j$ when starting at the state $\rho$.} \nonumber\\
& \tau(\rho\to j)=\text{expected time of first visit to $j$ when starting at the state $\rho$.\nonumber}
\end{flalign}
These notions are related with a given QMC $T$ and with the orthogonal projections onto the subspace generated by the vertices as follows. Let $\mathbb{P}_j$ be the projection matrix on vertex $j$ and let $\mathbb{Q}_j=\mathbb{I}-\mathbb{P}_j$ be its complement. Such projection is such that if $\rho$ is an OQW density then
$$\mathbb{P}_j\Big(\sum_i \rho_i\otimes|i\rangle\langle i|\Big)=\rho_j\otimes|j\rangle\langle j|$$
so we keep track to a visit to vertex $j$ regardless of the internal degree of freedom of the particle. Then we have:
\begin{flalign}
& p_r(\rho\to j)=\mathrm{Tr}(\mathbb{P}_jT^r\rho) \nonumber \\
& \pi_r(\rho\to j)=\mathrm{Tr}(\mathbb{P}_jT(\mathbb{Q}_jT)^{r-1}\rho) \nonumber \\
& \pi(\rho\to j)=\sum_{r\geq 1} \pi_r(\rho\to j)=\sum_{r\geq 1} \mathrm{Tr}(\mathbb{P}_jT(\mathbb{Q}_jT)^{r-1}\rho) \nonumber \\
 &\tau(\rho\to j)=\left\{
\begin{array}{ll}
\infty & \text{ if } \pi(\rho\to j)<1  \\
\sum_{r\geq 1} r\pi_r(\rho\to j) &  \text{ if } \pi(\rho\to j)=1 
\end{array} \right. \nonumber 
\end{flalign}
In the context of generalized open quantum walks the above notion of hitting time has been first discussed in \cite{gawron}, and we refer the reader to \cite{spsurvey} for more on the case of OQWs, noting that some of the works on recurrence mentioned in such survey follow a similar line. In order to obtain formulae associated with hitting times we will need to work with block matrices such that the individual blocks will lead to the trace calculations presented above. Having in mind technical developments of the theory, we introduce the following matrix-valued generating functions,
$$\mathbb{G}_{ij}(z)=\sum_{m\geq 1} \mathbb{P}_iT(\mathbb{Q}_i T)^{m-1}\mathbb{P}_jz^{m-1}=\mathbb{P}_iT(I-z\mathbb{Q}_iT)^{-1}\mathbb{P}_j,\;\;\;z\in\mathbb{D}$$
where $\mathbb{D}=\{z\in\mathbb{C}:|z|<1\}$. We remark that we will not explore the theory of such analytic functions, as these are related to the study of the so-called Schur functions and FR-functions associated with closed contractions in Banach spaces \cite{gv}. A systematic study of the hitting time versions of such objects will be the goal of a future work. Instead, in the present article we are focused on the more practical problem of calculating hitting times for QMCs on a finite graph. Regarding the above expressions, we are able to write
\begin{flalign}
 & \pi(\rho_j\to i)=\mathrm{Tr}(\hat{h}_{ij}\rho_j),\;\;\;\;\;\;\hat{h}_{ij}:=\left\{
\begin{array}{ll}
\lim_{x\uparrow 1}\mathbb{G}_{ij}(x) & \text{ if } i\neq j  \\
I &  \text{ if } i=j 
\end{array} \right. \\
 & \tau(\rho_i\to i)=\mathrm{Tr}(\hat{r}_{i}\rho_j),\;\;\;\;\;\;\hat{r}_{i}:=\lim_{x\uparrow 1}\frac{d}{dx}x\mathbb{G}_{ii}(x) \\
  &  \tau(\rho_j\to i)=\mathrm{Tr}(\hat{k}_{ij}\rho_j),\;\;\;\;\;\;\hat{k}_{ij}:=\lim_{x\uparrow 1}\frac{d}{dx}x\mathbb{G}_{ij}(x),\;\;\;\text{ if } i\neq j
\end{flalign}
noting that in such expressions the indices are read from right to left. For instance, $\hat{k}_{ij}$ is the transition operator from vertex $j$ to $i$.  We define by convenience $\hat{k}_{ii}:=0$, since the mean time of first visit is, by definition, a time $t\geq 0$ whereas the mean time of first {\it return} is assumed to be a time $t\geq 1$. Above we note that in order the calculate $\hat{k}_{ij}$, one makes use of the hitting probability kernel $\hat{h}_{ij}$ together with the usual derivation rule
$$\frac{d}{dz}(I-zA)^{-1}=(I-zA)^{-1}A(I-zA)^{-1},\;\;\;A\in M_k(\mathbb{C})$$
from which we obtain
$$\frac{d}{dx}x\mathbb{G}_{ij}(x)=\mathbb{G}_{ij}(x)+x\mathbb{G}_{ij}'(x)=\mathbb{P}_i\Phi(I-z\mathbb{Q}_i\Phi)^{-1}\mathbb{P}_j+x\mathbb{P}_i\Phi(I-x\mathbb{Q}_i\Phi)^{-1}\mathbb{Q}_i\Phi(I-x\mathbb{Q}_i\Phi)^{-1}\mathbb{P}_j$$
Finally, define
\beq\label{blockops}
H=\begin{bmatrix} \hat{h}_{11} & \cdots & \hat{h}_{1n} \\ \hat{h}_{21} & \cdots &  \hat{h}_{2n} \\ \vdots & \ddots & \vdots \\ \hat{h}_{n1} & \cdots & \hat{h}_{nn}\end{bmatrix},\;\;\;K=\begin{bmatrix} \hat{k}_{11} & \cdots & \hat{k}_{1n} \\ \hat{k}_{21} & \cdots &  \hat{k}_{2n} \\ \vdots & \ddots & \vdots \\ \hat{k}_{n1} & \cdots & \hat{k}_{nn}\end{bmatrix},\;\;\;D=\begin{bmatrix} \hat{r}_1 & 0 & \cdots & 0 \\ 0 & \hat{r}_2 & 0 & \cdots
\\ \vdots & \vdots & \cdots & \vdots \\
0 & 0 & \cdots & \hat{r}_n\end{bmatrix}\eeq
Such matrices of operators will play a central role in the description of hitting time formulae later in this work.

\section{How to write concrete calculations with QMCs}\label{sec3}

It is well-known that finite Markov chains with a countable state space can be studied in terms of its associated stochastic matrix, which on its turn allows for concrete calculations in terms of linear algebra, possibly with the use of a computer: given a stochastic matrix $P$, if $v=(v_i)$ is a probability vector then $P^nv$ gives the statistical distribution of the position of the walker at time $n$ (assuming $P$ is column stochastic). This practical aspect of stochastic matrices is certainly an attractive feature of Markov chains and a natural question is to ask whether one has algebraic tools in the case of QMCs which are as convenient as in the classical case. In this section we outline how this can be performed, while at the same time we illustrate similarities and differences with the case of stochastic matrices. In Section \ref{secex} we illustrate the constructions given here with examples.

\medskip

The main idea is to regard expression (\ref{defqmcgen}) as a matrix-vector computation. In order to do this, we need to consider the matrix representations of each $\hat{B}_{ij}$ or $\Phi_{ij}$ so that $T$ can be seen as a block matrix. At the same time, densities of the form $\rho=[\rho_1\;\cdots\;\rho_n]^T$ need to be written in vector form, as described below.

\medskip

If $A\in M_k(\mathbb{C})$, the corresponding vector representation $vec(A)$ associated to it is given by stacking together the matrix rows. For instance, if $n=2$,
$$A=\begin{bmatrix} a_{11} & a_{12} \\ a_{21} & a_{22}\end{bmatrix}\;\;\;\Rightarrow \;\;\; vec(A)=\begin{bmatrix} a_{11} & a_{12} & a_{21} & a_{22}\end{bmatrix}^T$$
The $vec$ mapping satisfies $vec(AXB^T)=(A\otimes B)vec(X)$ for any $A, B, X$ square matrices so, in particular, $vec(BXB^\dagger)=vec(BX\ov{B}^T)=(B\otimes \ov{B})vec(X)$
Then we can obtain the matrix representation for any CP map $\Phi$ \cite{hj2}:
$$\Phi(X):=\sum_i V_iXV_i^\dagger,\;\;\;[\Phi]:=\sum_{i} V_{i}\otimes \ov{V_{i}}\;\Longrightarrow\; \Phi(X)=vec^{-1}([\Phi]vec(X))$$
We will also use the notation $[A]:=A\otimes\ov{A}$ for any $A\in M_k(\mathbb{C})$. For a QMC density matrix, define
$$|\rho\rangle:=\begin{bmatrix} vec(\rho_1)\\ vec(\rho_2) \\ \vdots \\ vec(\rho_n)\end{bmatrix},\;\;\;\langle\rho|:=|\rho\rangle^\dagger,\;\;\;\rho=[\rho_1\cdots \rho_n]^T\in\mathcal{D}_{n;k}$$
Sometimes we will write $|A\rangle=vec(A)$ for general matrices $A\in M_n(\mathbb{C})$ as well. In this way, one iteration of the QMC $T$ given by (\ref{defqmcgen}) is equivalent to write
\beq\label{defqmcgen2} \rho\mapsto T(\rho)\;\Longleftrightarrow\;|\rho\rangle\mapsto \hat{T}|\rho\rangle=\begin{bmatrix} [\Phi_{11}] & \cdots & [\Phi_{1n}] \\ [\Phi_{21}] & \cdots &  [\Phi_{2n}] \\ \vdots & \ddots & \vdots \\ [\Phi_{n1}] & \cdots & [\Phi_{nn}]\end{bmatrix}\cdot\begin{bmatrix} vec(\rho_1) \\ vec(\rho_2) \\ \vdots \\ vec(\rho_n)\end{bmatrix},\;\;\;\rho\in\mathcal{D}_{n;k}\eeq
where the above expression is the usual matrix-vector multiplication. Higher iterates are obtained similarly. Above, for a QMC $T$ we use the notation $\hat{T}$ for what we call the {\it block matrix representation} for $T$. It is worth noting that $\hat{T}\neq [T]$, as their dimensions are distinct. If the QMC acts on $n$ vertices and the internal degree is $k$ then $[B_{ij}]$ has order $k^2$ so $\hat{T}$ is an order $nk^2$ matrix and since $\rho_i\in M_k(\mathbb{C})$, for all $i=1,\dots,n$, we have that $|\rho\rangle\in \mathbb{C}^{nk^2}$. From this we conclude that all spectral information of the QMC $T$ can be obtained by examining the block matrix $\hat{T}$.

\subsection{Irreducibility, aperiodicity and ergodicity}\label{secirred}

We recall that, by definition, a Markov chain is ergodic if it is irreducible, positive recurrent and aperiodic, and we refer the reader to \cite{bremaud} for these classical notions. Naturally we have the corresponding notions for CP maps. Since every positive trace preserving map on a finite dimensional space has an invariant state, we say that a QMC $T$ is {\it irreducible} if  $T$ admits a unique invariant state $\pi=\sum_i \pi_i\otimes|i\rangle\langle i|$, $\pi_i\in M_k(\mathbb{C})$, and if such $\pi$ is strictly positive, i.e., $\pi_i>0$ for every $i$. As we will only consider irreducible CP maps acting on finite dimensional spaces, we say a QMC $T$ on a finite graph is {\it aperiodic} if 1 is the only eigenvalue of $T$ with unit  modulus. Then, we say that a finite QMC is {\it ergodic} if it is  irreducible and aperiodic. It is well-known that the iterates of an ergodic QMC acting on any initial density converge to $\pi$ \cite{cp}. We remark that in \cite{cp} the term ergodic refers to a distinct notion than the one employed here.

\subsection{The iterates of an ergodic chain}

If $P$ is the column stochastic matrix associated with an ergodic chain, we know that its powers converge to a limit matrix for which its columns are all equal to the (unique) stationary vector $\pi$, that is,
\beq\label{ergodiclim} P^m\to \Pi=\begin{bmatrix} \pi_1 & \pi_1 & \cdots & \pi_1 \\ \pi_2 & \pi_2 & \cdots & \pi_2 \\ \vdots & \vdots & \cdots & \vdots \\ \pi_n & \pi_n & \cdots & \pi_n\end{bmatrix}=|\pi\rangle\langle 1|,\;\;\;|\pi\rangle=\begin{bmatrix} \pi_1 \\ \vdots\\ \pi_n\end{bmatrix},\;\;\;\langle 1|:=\begin{bmatrix} 1  & 1 & \cdots & 1\end{bmatrix}\eeq
as $m\to\infty$. On the other hand, we may consider the class of finite ergodic QMCs and then inspect the behavior of the associated block  matrix representation $\hat{T}$. A moment's thought makes clear that the statement analogous to (\ref{ergodiclim}), i.e., that as $m\to\infty$, $\hat{T}^m$ converges to a matrix for which all columns are equal, is not true in general (unless $k$, the internal degree of freedom, equals 1). Instead, what happens is that the limit QMC is of the form
\beq\hat{T}^m\to \hat{\Omega}=|\pi\rangle\langle e_{I_k^n}|,\;\;\;\;\;\;|\pi\rangle, |e_{I_k^n}\rangle:=\begin{bmatrix} vec(I_k) \\ vec(I_k) \\ \vdots \\ vec(I_k)\end{bmatrix}\in\mathbb{C}^{nk^2}\eeq
as $m\to\infty$, where $\pi$ is the unique stationary state for $T$ and $I_k\in M_k(\mathbb{C})$ is the order $k$ identity matrix. For instance, if $n=k=2$, write
$$\pi=\pi_1\otimes|1\rangle\langle 1|+\pi_2\otimes|2\rangle\langle 2|=\begin{bmatrix} \pi_{11} & \pi_{12} \\ \pi_{21} & \pi_{22}\end{bmatrix}\otimes|1\rangle\langle 1|+\begin{bmatrix} \pi_{33} & \pi_{34} \\ \pi_{43} & \pi_{44}\end{bmatrix}\otimes|2\rangle\langle 2|,\;\;\;|\pi\rangle=\begin{bmatrix} vec(\pi_1) \\ vec(\pi_2)\end{bmatrix}$$
so the limit matrix has the form
$$\hat{\Omega}=|\pi\rangle\langle e_{I_2^2}|=\begin{bmatrix} \pi_{11} \\ \pi_{12} \\ \pi_{21} \\ \pi_{22} \\ \pi_{33} \\ \pi_{34} \\ \pi_{43} \\ \pi_{44} \end{bmatrix}\begin{bmatrix} 1 & 0 & 0 & 1 & 1 & 0 & 0 & 1\end{bmatrix}=\begin{bmatrix} 
\pi_{11} & 0 & 0 & \pi_{11} & \pi_{11} & 0 & 0 & \pi_{11}\\
\pi_{12} & 0 & 0 & \pi_{12} & \pi_{12} & 0 & 0 & \pi_{12}\\
\pi_{21} & 0 & 0 & \pi_{21} & \pi_{21} & 0 & 0 & \pi_{21}\\
\pi_{22} & 0 & 0 & \pi_{22} & \pi_{22} & 0 & 0 & \pi_{22}\\
\pi_{33} & 0 & 0 & \pi_{33} & \pi_{33} & 0 & 0 & \pi_{33}\\
\pi_{34} & 0 & 0 & \pi_{34} & \pi_{34} & 0 & 0 & \pi_{34}\\
\pi_{43} & 0 & 0 & \pi_{43} & \pi_{43} & 0 & 0 & \pi_{43}\\
\pi_{44} & 0 & 0 & \pi_{44} & \pi_{44} & 0 & 0 & \pi_{44}
\end{bmatrix}$$
For simplicity we use the notation $|e_{I_k^n}\rangle=|e_I\rangle$ in case $n$ and $k$ are clear from context.

\subsection{Conditioning on the first step}

We recall the following classical reasoning. For a walk starting at vertex $j$, consider the mean time of first visit to vertex $i$, $i\neq j$: take the expected number of steps required given the outcome
of the first step, multiply by the probability that this outcome occurs, and add. If the first step is to $i$, the expected number of steps required is 1 and if it is to some other vertex, say $l$, the expected number of steps required is $k_{il}$ plus 1 for the step already taken (recall indices for $k$ are read from right to left). Therefore,
$$k_{ij}=p_{ij}+\sum_{l\neq i} (k_{il}+1)p_{lj}=1+\sum_{l\neq i} k_{il}p_{lj}$$
$$r_i=1+\sum_l k_{il}p_{li}$$
where $r_i$ denotes the {\it mean time of first return} to vertex $i$. This can be written in matrix form: for instance,   in the case $n=3$, we have
$$\begin{bmatrix} r_1 & 0 & 0 \\ 0 & r_2 & 0 \\ 0 & 0 & r_3\end{bmatrix}+\begin{bmatrix} 0 & k_{12} & k_{13}\\ k_{21} & 0 & k_{23} \\ k_{31} & k_{32} & 0\end{bmatrix}=\begin{bmatrix} 1 & 1 & 1 \\ 1 & 1 & 1 \\ 1 & 1 & 1\end{bmatrix}+\begin{bmatrix} 0 & k_{12} & k_{13} \\ k_{21} & 0 & k_{23} \\ k_{31} & k_{32} & 0\end{bmatrix}\begin{bmatrix} p_{11} & p_{12} & p_{13} \\ p_{21} & p_{22} & p_{23} \\ p_{31} & p_{32} & p_{33}\end{bmatrix}$$
which we write as
$$K=D+(K-D)=E+(K-D)P$$
where $E$ denotes the matrix with all entries equal to 1 and $D$ denotes the  diagonal matrix with nonzero entries equal to the mean return time. We will write the above matrix equation, characterizing the conditioning on the first step for the classical case, as
\beq\label{cfsclassic} E=K-(K-D)P\eeq
Now we turn to the case of QMCs. For $\rho_j$ density concentrated on vertex $j$ we obtain in a similar way as in the classical case that \cite{oqwmhtf}
$$k_{ij}(\rho_j)=1+\sum_{l\neq i}k_{il}\Big(\frac{B_{lj}\rho_j B_{lj}^\dagger }{\mathrm{Tr}(B_{lj}\rho_j B_{lj}^\dagger )}\Big)\mathrm{Tr}(B_{lj}\rho_j B_{lj}^\dagger )
$$
where traces appear on the right with the purpose of making explicit the probabilistic reasoning. By simplifying the trace, 
$$
k_{ij}(\rho_j)=1+\sum_{l\neq i}k_{il}(B_{lj}\rho_j B_{lj}^\dagger )\;\Longrightarrow\; k_{ij}(\rho_j)-\sum_{l\neq i}k_{il}(B_{lj}\rho_j B_{lj}^\dagger )=1$$
Now we note that we are able to write a block matrix expression in analogy to expression (\ref{cfsclassic}). We replace the stochastic matrix $P$ with the CP map $T$ describing the QMC and by considering the mean hitting time operator (\ref{blockops}), we may define
\beq L:=K-(K-D)T\eeq
In the classical case we know that $L$ equals matrix $E$. However, in the QMC case, $L$ (i.e., its matrix representation) does not have all entries equal to 1 in general. Nevertheless, we have the important fact, proved in \cite{oqwmhtf}, that for every density $\rho_j$ concentrated on a vertex $j$, for all $i$,
\beq\label{lleq1} \mathrm{Tr}(\hat{L}_{ij}\rho_j)=1\eeq
where $\hat{L}_{ij}$ is the operator corresponding to the $(i,j)$-th block matrix representation appearing in $\hat{L}$. If the internal degree of freedom equals 1 then (\ref{lleq1}) means simply that $E_{ij}=1$. Such fact will play an essential role in the proof of the results on generalized inverses that follows.

\section{Generalized inverses}\label{sec4}

{\it Definition.} A generalized inverse, or {\it $g$-inverse}, of a matrix $A$ is any matrix $A^-$ such that
\beq\label{ginvdef}
AA^-A=A\eeq

\medskip

Generalized inverses are in general not unique (unless $A$ is nonsingular, in which case $A^-=A^{-1}$). As it is well-known, by imposing additional conditions one may obtain a unique generalized inverse.

\medskip

Let $A\{1\}$ denote the set of one condition $g$-inverses $A^-$ of $A$ satisfying equation (\ref{ginvdef}). The following results are basic to our applications and can be applied immediately to our context since these are just linear algebraic results. The proofs of the other results in this section are given in the Appendix, and are motivated by the classical results seen in \cite{hunter}.

\begin{pro}\cite{campbell,hunter} If $A^-$ is any $g$-inverse of $A$ then all $g$-inverses can be described as members of the following equivalent sets:
\beq\label{eq38}A\{1\}=\{A^-+H-A^-AHAA^-:H \text{ arbitrary}\}\eeq
\beq\label{eq39}=\{A^-+(I-A^-A)F+G(I-AA^-): F,G \text{ arbitrary}\}\eeq
\end{pro}

\begin{pro}\cite{campbell,hunter}\label{corsol1}
a) A necessary and sufficient condition for the equation $AX=C$ to be consistent is that $AA^-C=C$, where $A^-$ is any $g$-inverse of $A$, in which case the general solution is given by 
$$X=A^-C+(I-A^-A)U$$
where $U$ is an arbitrary matrix. b) A necessary and sufficient condition for the equation $XB=C$ to be consistent is that $CB^-B=C$, where $B^-$ is any $g$-inverse of $B$, in which case the general solution is given by 
$$X=CB^-+V(I-BB^-)$$
where $V$ is an arbitrary matrix.
\end{pro}

\medskip

The existence of $g$-inverses associated with QMCs is established by the following result. Note that we can obtain such inverses even without the assumption of ergodicity.

\begin{pro}\label{hunter33} Let $T $ be an irreducible QMC on a finite graph with stationary density $\pi$. Let $|t\rangle, | u\rangle\in\mathbb{C}^n$ be such that $\langle e_I|t\rangle\neq 0$ and $\langle u|\pi\rangle\neq 0$. Then $I-T +|t\rangle\langle u|$ is invertible and its inverse is a $g$-inverse of $I-T $.
\end{pro} 
	
\begin{cor}\label{giexps} Under the conditions of Proposition \ref{hunter33}, any $g$-inverse $G_0$ of $I-T$ can be written in one of the following forms:
\begin{flalign}
 & a) \; G_0=(I-T+|t\rangle\langle u|)^{-1}+H\frac{|t\rangle\langle e_I|}{\langle e_I|t\rangle}+\frac{|\pi\rangle\langle u|}{\langle u|\pi\rangle}H-\frac{|\pi\rangle\langle u|H|t\rangle\langle e_I|}{\langle u|\pi\rangle\langle e_I|t\rangle}  \nonumber \\
 & b) \; G_0=(I-T+|t\rangle\langle u|)^{-1}+\frac{|\pi\rangle\langle u|}{\langle u|\pi\rangle}F+G\frac{|t\rangle\langle e_I|}{\langle e_I|t\rangle} \nonumber \\
 & c)\; G_0=(I-T+|t\rangle\langle u|)^{-1}+|\pi\rangle\langle f|+|g\rangle\langle e_I| \nonumber
 \end{flalign}
where $f,g$ are arbitrary vectors, $F,G,H$ are arbitrary matrices.
\end{cor}

\medskip

Set $\langle f|$, $|g\rangle$ to be the null vector, $|t\rangle=|\pi\rangle$ and $\langle u|=\langle e_I|$, so we immediately obtain:

\begin{cor}\label{corzz} Let $T$ be an irreducible QMC on a finite graph with stationary density $\pi$ and let $\Omega=|\pi\rangle\langle e_I|$. Then 
\beq\label{zqmc} Z=(I-T+\Omega)^{-1}\eeq
is a generalized inverse of $I-T$.\end{cor}

\medskip

The map $Z$ defined above is called the {\it fundamental matrix} associated with the QMC $T$. In Section \ref{sec5} we will note its usefulness in the problem of hitting times for both classical and quantum Markov chains. In the classical setting the origin of $Z$ comes from the notion of a potential matrix \cite{bremaud},
$$W=\sum_{m=0}^\infty T^m$$
so each entry of $W$ counts the mean number of visits to a particular vertex given some initial position, noting that  we only consider pairs of vertices $(i,j)$ for which $i,j$ are transient (in the other situations we obtain  null or infinite entries). On the other hand, it turns out that in the irreducible case we can make a proper modification of $W$, namely replace $T^n$ with $T^n-\Omega$ from which we obtain (\ref{zqmc}).

\subsection{Stationary densities in terms of generalized inverses}

The following result on fixed points may be of independent interest and we illustrate its use in the examples.

\begin{pro}\label{teo41} Let $T$ be an irreducible QMC on a finite graph. a) If $(I-T)^-$ is any $g$-inverse of $I-T$ and if $A=I-(I-T)^-(I-T)$ then
\beq\label{fixeq41}
|\pi\rangle=\frac{A|v\rangle}{\langle e_I |Av\rangle}\eeq
where $|v\rangle$ is any vector such that $\langle e_I |Av\rangle\neq 0$.
b) If we set
$$(I-T)^-=(I-T+|t\rangle\langle u|)^{-1}+|f\rangle\langle e_I|$$
where $|f\rangle$ is arbitrary and $|t\rangle$, $\langle u|$ are such that $\langle e_I|t\rangle\neq 0$ and $\langle u|\pi\rangle\neq 0$ then
\beq\label{fixformula}
|\pi\rangle=\frac{(I-T +|t\rangle\langle u|)^{-1}|t\rangle}{\langle e_I |(I-T +|t\rangle\langle u|)^{-1}|t\rangle}\eeq
\end{pro}

\subsection{Hitting times in terms of generalized inverses}

The following is one of our main results. The proof is inspired by the result from \cite{hunter} on Markov chains, relies on the block matrix representation of the QMC and on the matrix expressions described in Section \ref{sec3}. If $B=(\hat{b}_{ij})$ is a matrix of operators, we use the notation $B_d:=diag(\hat{b}_{11},\dots,\hat{b}_{nn})$. 

\begin{maintheorem}\label{bigtheorem} (Hunter's formula for QMCs). Let $T$ be an ergodic QMC on a finite graph with $n\geq 2$ vertices and internal degree $k\geq 2$, let $\pi$ be its stationary density and $\Omega$ its limit map. Let $K=(\hat{k}_{ij})$ denote the matrix of mean hitting time operators to vertices $i=1,\dots,k$, $D=K_d=diag(\hat{k}_{11},\dots,\hat{k}_{nn})$, $G$ be any $g$-inverse of $I-T$ and let $E$ denote the block matrix for which each block equals the identity of order $k^2$. a) The mean hitting time for the walk to reach vertex $i$, beginning at vertex $j$ with initial density $\rho_j$ is given by
\beq\label{hunterf1}
\tau(\rho_j\otimes|j\rangle\to |i\rangle)=\mathrm{Tr}(\hat{k}_{ij}\rho_j)=\mathrm{Tr}\Big(\Big[D\Big(\Omega G -(\Omega G)_dE+I-G+G_dE\Big)\Big]_{ij}\rho_j\Big)\eeq
b) By setting $G=(I-T+|u\rangle\langle e_{I}|)^{-1}+|f\rangle\langle e_I|$, with $|f\rangle$ arbitrary and $|u\rangle$ such that $\langle u|\pi\rangle\neq 0$, we have that for every vertex $i$ and initial density $\rho_j$ on vertex $j$,
\beq\label{hunterf2}
\tau(\rho_j\otimes|j\rangle\to |i\rangle)=\mathrm{Tr}(\hat{k}_{ij}\rho_j)=\mathrm{Tr}\Big(\Big[D\Big(I-G+G_dE\Big)\Big]_{ij}\rho_j\Big)\eeq
\end{maintheorem}
Informally, the meaning of the theorem is: the mean time of first visit from $j$ to $i$ is an information which can be extracted from the mean {\it return} time to vertices if we have knowledge of any generalized inverse of $I-T$. This particular aspect of the above formulae will appear again in the following section.

\section{The fundamental matrix and hitting time formulae}\label{sec5}

If $T$ is an ergodic QMC with limit map $\Omega=|\pi\rangle\langle e_I|$, let $Z$ denote its fundamental matrix as defined by (\ref{zqmc}) and denote by $\hat{ Z}_{ij}$ the operator in the $(i,j)$-th block of $ \hat{Z}$, the block  matrix representation of $Z$. We have the following:

\begin{theorem}\label{oqwmhtfstat}(First mean hitting time formula). Let $T$ be an ergodic QMC acting on a finite graph with $n\geq 2$ vertices and let $D$ denote the block diagonal matrix with block diagonal entries given by the operators $\hat{k}_{ii}$, $i=1,\dots,n$. a) The mean hitting time for the walk to reach vertex $i$, beginning at vertex $j$ with initial density $\rho_j$ concentrated in vertex $j$ is given by
\beq\label{aldous_eqnova0}
\tau(\rho_j\otimes|j\rangle\to |i\rangle)=\mathrm{Tr}(\hat{k}_{ij}\rho_j)=\mathrm{Tr}(\hat{k}_{ii}(\hat{ Z}_{ii}-\hat{ Z}_{ij})\rho_j)\eeq
b) (Random target lemma). If  $D$ is invertible and there is $c$ scalar such that $\mathrm{Tr}(\hat{k}_{ii}\gamma)=c\mathrm{Tr}(\gamma)$, all $i$ vertex, $\gamma\in M_n(\mathbb{C})$, then for every density $\rho$,
\begin{equation}\label{aldous_eqnova}
\mathrm{Tr}[({D}^{-1}K)_{ij}\rho]=\mathrm{Tr}[({\hat{ Z}}_{ii}-{\hat{ Z}}_{ij})\rho]
\end{equation}
As a consequence,
\begin{equation}\label{rtlem11}
t_{\odot}(\rho):=\sum_i \mathrm{Tr}[({D}^{-1}{K})_{ij}\rho]=\Big[\sum_i \mathrm{Tr}(\hat{Z}_{ii}\rho)\Big]-1
\end{equation}
In particular, such quantity does not depend on $j$.

\end{theorem}

Equation (\ref{aldous_eqnova0}) is to be regarded as a generalization of the classical expression (\ref{clmhtf}). The fact that such block matrix version for the hitting time formula is valid is one of the main motivations for attempting further developments in the study of hitting times for QMCs. In a similar way as in Theorem \ref{bigtheorem}, we remark on the meaning of the formula (\ref{aldous_eqnova0}), namely the mean hitting time from $j$ to $i$ is an information which can be extracted from the mean return time to $i$ if we know the fundamental matrix of the walk.

\medskip

In addition, expression (\ref{rtlem11}) should be compared with the classical fact that if $T_i$ denotes the time of first visit to $i$ then the expected length of the walk with respect to a randomly chosen target $i$, 
$$
\sum_i\pi_i E_j(T_i)
$$
does not depend on the initial vertex $j$ \cite{aldous}. From this we see that the density matrix degree of freedom may modify the classical case in a nontrivial way. We call the number $t_{\odot}(\rho)$ the {\it target time} of the QMC with respect to the initial density $\rho$.

\medskip

Theorem \ref{oqwmhtfstat} is a refinement of the result appearing in \cite{oqwmhtf}, which is given in terms of unital OQWs. It is clear that the result holds in the nonunital case as well, a fact which follows from our  discussion on the form of $\Omega$, the limit matrix of an ergodic QMC. In addition, the proof actually holds for any finite ergodic QMC, as the reasoning relies exclusively in the block matrix structure of such objects, for which OQWs are a particular subset of channels (because of this we regard such class as being the key example). We refer the reader to \cite{oqwmhtf} for the proof of \ref{oqwmhtfstat}(a), and also to the examples in Section \ref{secex}. The proof of \ref{oqwmhtfstat}(b) is given in the Appendix.

\medskip

Now let
\beq F_{j\pi}:=\sum_i \hat{h}_{ji}(\pi_i),\;\;\;K_{j\pi}:=\sum_i \hat{k}_{ji}(\pi_i)\eeq
so $\mathrm{Tr}(F_{j\pi})$ is the probability of ever reaching vertex $j$ given some initial vertex randomly chosen according to the stationary density $\pi$ and $\mathrm{Tr}(K_{j\pi})$ is the mean hitting time for vertex $j$, given an initial density randomly chosen according to  $\pi$. In our context a random choice according to $\pi$ means that a vertex $i$ is chosen with probability $\mathrm{Tr}(\pi_i)$ and in this case the state is prepared to be $(\pi_i/\mathrm{Tr}(\pi_i))\otimes|i\rangle\langle i|$. The above definitions are naturally associated with the classical probabilistic notions
$$ P_\pi(X=j)=\sum_i \pi_iP_i(X=j),\;\;\;\;\;\;E_\pi(X)=\sum_x xP_\pi(X=x)=\sum_i\pi_i\sum_x xP_i(X=x)$$
where $X$ is a discrete random variable. We prove:

\begin{theorem}\label{MHTF2}(Second mean hitting time formula). Let $T$ be an ergodic QMC acting on a finite graph with $n\geq 2$ vertices, let $\hat{Z}$ denote its fundamental matrix and $\pi=(\pi_i)$ the unique stationary density for $T $. Let $D$ be the block diagonal matrix with block diagonal entries equal to the operators $\hat{k}_{ii}$. Then $\mathrm{Tr}(K_{j\pi})$ equals the mean hitting time for vertex $j$, given an initial density randomly chosen according to  $\pi$, and is given by
\beq\label{mhtf2eq}
\mathrm{Tr}(K_{j\pi})=\mathrm{Tr}[(D Z)_{jj}F_{j\pi}]\eeq
\end{theorem}

Equation (\ref{mhtf2eq}) is to be regarded as a quantum version of the classical result \cite{aldous,peres}
$$E_{\pi}(T_j)=\frac{Z_{jj}}{\pi_j}$$
In words: the mean time of first visit to $j$, given that it has started at a vertex randomly chosen according to $\pi$, is an information dependent on the $j$-th diagonal position (block) of the fundamental matrix. The proof of the formula makes clear the appearance of the operator $F_{j\pi}$, and the reason why this term does not appear in the classical expression is simple: in this case the operator equals 1.

\section{Examples}\label{secex}

\bex\label{exemplo1}
What is the simplest nonclassical QMC on two vertices? We assume the internal degree of freedom equals $k\geq 2$, so we seek transition effects $B_{ij}$ of order $k$. Then, for instance, we can take appropriate multiples of the identity, but this is a trivial example which essentially reduces to the classical case. A less trivial example is given by restricting to $B_{ij}$ which are permutations of diagonal matrices, sometimes called PQ-matrices \cite{oqwmhtf}. The dynamics is relatively simple to study and the statistics obtained still resemble a classical stochastic matrix. A distinct example is the following: by applying a decoherence procedure to a unitary evolution for a coined unitary walk (see e.g. the realization procedure for an OQW in \cite{attal}) we are led to consider the following effects, induced by a unitary coin,
$$B_{11}=B_{22}=B=\begin{bmatrix} a & \sqrt{1-a^2} \\ 0 & 0 \end{bmatrix},\;\;\;B_{12}=B_{21}=C=\begin{bmatrix} 0 & 0 \\
-\sqrt{1-a^2} & a\end{bmatrix},\;\;\;0<a<1$$
and we may define the following OQW (see Figure \ref{fig1}):
$$
\hat{T}:=\begin{bmatrix} [B_{11}] & [B_{12}] \\ [B_{21}] & [B_{22}]\end{bmatrix}=\begin{bmatrix} 
a^2 & ab & ab & b^2 & 0 & 0 & 0 & 0 \\
0 & 0 & 0 & 0 & 0 & 0 & 0 & 0 \\
0 & 0 & 0 & 0 & 0 & 0 & 0 & 0 \\
0 & 0 & 0 & 0 & b^2 & -ab & -ab & a^2 \\
0 & 0 & 0 & 0 & a^2 & ab & ab & b^2 \\
0 & 0 & 0 & 0 & 0 & 0 & 0 & 0 \\
0 & 0 & 0 & 0 & 0 & 0 & 0 & 0 \\
b^2 & -ab & -ab & a^2 & 0 & 0 & 0 & 0 
\end{bmatrix},\;\;\;b:=\sqrt{1-a^2}$$

\begin{center}
\begin{figure}[ht]
\begin{tikzpicture}
[->,>=stealth',shorten >=1pt,auto,node distance=2.0cm,
                    semithick]
    \node[main node] (1) {$1$};
    \node[main node] (2) [right = 2.0cm and 2.0cm of 1]  {$2$};


    \path[draw,thick]
    (1) edge   [loop left]     node {$B$} (1)


    (1) edge    [bend left] node[above] {$C$}    (2)
    (2) edge   [bend left]     node[below] {$C$} (1)

    (2) edge   [loop right]     node {$B$ } (2)







    ;
\end{tikzpicture}
\caption{Graph with 2 vertices. In the block matrix $\widehat{T}$ we have that the $(i,j)$-th order 4 block matrix  equals the matrix representation $[B_{ij}]=B_{ij}\otimes\ov{B_{ij}}$ for the conjugation $B_{ij} \cdot B_{ij}^\dagger=C\cdot C^\dagger$ or $B\cdot B^\dagger$ which describes the possible transitions on such graph ($B_{ij}$ corresponds to the transition effect from vertex $j$ to vertex $i$). The choice of matrices given for such example leads to an ergodic, unital OQW.}\label{fig1}
\end{figure}
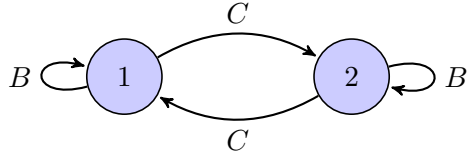
\end{center}

The eigenvalues of $\hat{T}$ are $1, 2a^2-1, \pm\sqrt{2a^2-1}$, and the fixed density $\pi$ equals $1/4$ times  the identity, as this can be verified with formula (\ref{fixformula}) (take for instance $|t\rangle=|u\rangle=[1\; 0 \; 0\; 0 \;0 \;0 \;0 \;0]^T$) so such OQW is ergodic and unital. Also,
$$\hat{\Omega}=|\pi\rangle\langle e_I|=\frac{1}{4}\begin{bmatrix} 1 \\ 0 \\0 \\ 1\\ 1 \\ 0 \\ 0 \\ 1\end{bmatrix}\begin{bmatrix} 1 & 0 & 0 & 1 & 1 & 0 & 0 & 1\end{bmatrix}=\begin{bmatrix} \Omega_{11} & \Omega_{11} \\ \Omega_{11} & \Omega_{11}\end{bmatrix},\;\;\;\Omega_{11}=\frac{1}{4}\begin{bmatrix} 1 & 0 & 0 & 1 \\ 0 & 0 & 0 & 0 \\ 0 & 0 & 0 & 0 \\ 1 & 0 & 0 & 1 \end{bmatrix}$$
$$\hat{h}_{11}=\hat{h}_{22}=\begin{bmatrix} 
a^2 & ab & ab & b^2\\
0 & 0 & 0 & 0 \\
0 & 0 & 0 & 0 \\
b^2 & -ab & -ab & a^2\end{bmatrix},\;\;\;\hat{h}_{12}=\hat{h}_{21}=\begin{bmatrix}
 0 & 0 & 0 & 0\\
 0 & 0 & 0 & 0\\
 0 &0 & 0 & 0 \\
 1 & 0 & 0 & 1 \end{bmatrix}$$
$$\hat{k}_{11}=\hat{k}_{22}=\begin{bmatrix} 
a^2 & ab & ab & b^2\\
0 & 0 & 0 & 0 \\
0 & 0 & 0 & 0 \\
3b^2 & -3ab & -3ab & 3a^2\end{bmatrix},\;\;\;\hat{k}_{12}=\hat{k}_{21}=\begin{bmatrix}
0 & 0 & 0 & 0 \\ 0 & 0 & 0 & 0 \\ 0 & 0 & 0 & 0 \\
\frac{1}{b^2} & \frac{a}{b} & \frac{a}{b} & 2\end{bmatrix}$$
From this we obtain that for every density and vertex the hitting probability equals 1, as expected, since this QMC is irreducible, and
$$\mathrm{Tr}(\hat{k}_{11}\rho)=\mathrm{Tr}(\hat{k}_{22}\rho)=(3-2a^2)\rho_{11}+(1+2a^2)\rho_{22}-4ab Re(\rho_{12}) $$
\beq\label{mhtex1}\mathrm{Tr}(\hat{k}_{12}\rho)=\mathrm{Tr}(\hat{k}_{21}\rho)=\frac{1}{b^2}\rho_{11}+2\rho_{22}+\frac{2a}{b}Re(\rho_{12})\eeq
The block matrix representation of the fundamental matrix is the order 8 matrix
$$\hat{Z}=(\hat{I}-\hat{T}+\hat{\Omega})^{-1}=\begin{bmatrix} \hat{Z}_{11} & \hat{Z}_{12} \\ \hat{Z}_{21} & \hat{Z}_{22}\end{bmatrix}= \begin{bmatrix}
\frac{5}{8b^2} & \frac{3a}{4b} & \frac{3a}{4b} & -\frac{4a^2-3}{8b^2} & -\frac{4a^2-1}{8b^2} & -\frac{a}{4b} & -\frac{a}{4b} & -\frac{1}{8b^2}\\
0 & 1 & 0 & 0 & 0 & 0 & 0 & 0 \\
0 & 0 & 1 & 0 & 0 & 0 & 0 & 0 \\
-\frac{1}{8b^2} & -\frac{a}{4b} & -\frac{a}{4b} & -\frac{4a^2-5}{8b^2} & -\frac{4a^2-3}{8b^2} & -\frac{a}{4b} & -\frac{a}{4b} & \frac{1}{8b^2}\\
-\frac{4a^2-1}{8b^2} & -\frac{a}{4b} & -\frac{a}{4b} & -\frac{1}{8b^2} & \frac{5}{8b^2} & \frac{3a}{4b} & \frac{3a}{4b} & -\frac{4a^2-3}{8b^2} \\
0 & 0 & 0 & 0 & 0 & 1 & 0 & 0\\
0 & 0 & 0 & 0 & 0 & 0 & 1 & 0 \\
-\frac{4a^2-3}{8b^2} & -\frac{a}{4b} & -\frac{a}{4b} & \frac{1}{8b^2} & -\frac{1}{8b^2} & -\frac{a}{4b} & -\frac{a}{4b} & -\frac{4a^2-5}{8b^2}
\end{bmatrix}$$
We can apply $Z$ to Theorem \ref{oqwmhtfstat}, so
$$\hat{k}_{11}(\hat{ Z}_{11}-\hat{ Z}_{12})=\begin{bmatrix} 
a^2 & ab & ab & b^2\\
0 & 0 & 0 & 0 \\
0 & 0 & 0 & 0 \\
3b^2 & -3ab & -3ab & 3a^2\end{bmatrix}\begin{bmatrix} \frac{1+a^2}{2b^2} & \frac{a}{b} & \frac{a}{b} & \frac{1}{2} \\
0 & 1 & 0 & 0 \\
0 & 0 & 1 & 0 \\
-\frac{1}{2} & 0 & 0 & \frac{1}{2}\end{bmatrix}=\begin{bmatrix} \frac{3a^2-1}{2b^2} & \frac{a}{b} & \frac{a}{b} & \frac{1}{2}\\
0 & 0 & 0 & 0 \\
0 & 0 & 0 & 0 \\
\frac{3}{2} & 0 & 0 & \frac{3}{2}\end{bmatrix}
$$
Therefore for $\rho=(\rho_{ij})_{i,j=1,2}$ density matrix concentrated on vertex 2,
$$vec^{-1}[\hat{k}_{11}(\hat{ Z}_{11}-\hat{ Z}_{12})vec(\rho)]=\begin{bmatrix} \frac{3a^2-1}{2b^2}\rho_{11} +\frac{\rho_{22}}{2}+\frac{2a}{b}Re(\rho_{12}) & 0 \\ 0 & \frac{3}{2}(\rho_{11}+\rho_{22})\end{bmatrix}$$
from which we obtain (\ref{mhtex1}), as expected. In a similar way Hunter's formula can be verified and we illustrate this by making a particular choice of $g$-inverse. Take $|u\rangle$ as before and define 
$$G=(I-T+|u\rangle\langle e_{I}|)^{-1}=\begin{bmatrix} \hat{G}_{11} & \hat{G}_{12} \\ \hat{G}_{21} & \hat{G}_{22}\end{bmatrix}=\frac{1}{4}\begin{bmatrix}
5 & \frac{3a}{b} & \frac{3a}{b} & -\frac{7a^2-4}{b^2} & -\frac{7a^2-3}{b^2} & -\frac{a}{b} & -\frac{a}{b} & -\frac{5a^2-2}{b^2}\\
0 & 4 & 0 & 0 & 0 & 0 & 0 & 0 \\
0 & 0 & 4 & 0 & 0 & 0 & 0 & 0 \\
1 & -\frac{a}{b} & -\frac{a}{b} & -\frac{3a^2-4}{b^2} & 3 & -\frac{a}{b} & -\frac{a}{b} & -\frac{a^2-2}{b^2}\\
1 & -\frac{a}{b} & -\frac{a}{b} & \frac{a^2}{b^2} & \frac{a^2+3}{b^2} & \frac{3a}{b} & \frac{3a}{b} & -\frac{a^2-2}{b^2}\\
0 & 0 & 0 & 0 & 0 & 4 & 0 & 0 \\
0 & 0 & 0 & 0 & 0 & 0 & 4 & 0 \\
1 & -\frac{a}{b} & -\frac{a}{b} & \frac{a^2}{b^2} & -1 & -\frac{a}{b} & -\frac{a}{b} & -\frac{a^2-2}{b^2}
\end{bmatrix}
$$
We set $G_d=diag(\hat{G}_{11},\hat{G}_{22})$ so
$$D(I-G+G_dE)$$
$$=\begin{bmatrix}
a^2 & ab & ab & b^2 & 0 & 0 & 0 & 0 \\
0 & 0 & 0 & 0 & 0 & 0 & 0 & 0 \\
0 & 0 & 0 & 0 & 0 & 0 & 0 & 0 \\
3b^2 & -3ab & -3ab & 3a^2 & 0 & 0 & 0 & 0 \\
0 & 0 & 0 & 0 & a^2 & ab & ab & b^2\\
0 & 0 & 0 & 0 & 0 & 0 & 0 & 0 \\
0 & 0  & 0 & 0 & 0 & 0 & 0 & 0 \\
0 & 0 & 0 & 0 & 3b^2 & -3ab & -3ab & 3a^2
\end{bmatrix}
\begin{bmatrix} 
1 & 0 & 0 & 0 & \frac{a^2+1}{2b^2} & \frac{a}{b} & \frac{a}{b} & \frac{1}{2}\\
0 & 1 & 0 & 0 & 0 & 1 & 0 & 0 \\
0 & 0 & 1 & 0 & 0 & 0 & 1 & 0 \\
0 & 0 & 0 & 1 & -\frac{1}{2} & 0 & 0 & \frac{1}{2}\\
\frac{a^2+1}{2b^2} & \frac{a}{b} & \frac{a}{b} & \frac{1}{2} & 1 & 0 & 0 &  0\\
0 & 1 & 0 & 0 & 0 & 1 & 0 & 0 \\
0 &  0 & 1 & 0 & 0 & 0 & 1 & 0 \\
-\frac{1}{2} & 0 & 0 & \frac{1}{2} & 0 & 0 & 0 & 1
\end{bmatrix}$$
$$=\begin{bmatrix}
a^2 & ab & ab & b^2 & \frac{3a^2-1}{2b^2} & \frac{a}{b} & \frac{a}{b} & \frac{1}{2}\\
0 & 0 & 0 & 0 & 0 & 0 & 0 & 0 \\
0 & 0 & 0 & 0 & 0 & 0 & 0 & 0 \\
3-3a^2 & -3ab & -3ab & 3a^2 & \frac{3}{2} & 0 & 0 & \frac{3}{2}\\
\frac{3a^2-1}{2b^2} & \frac{a}{b} & \frac{a}{b} & \frac{1}{2} & a^2 & ab & ab & b^2 \\
0 & 0 & 0 & 0 & 0 & 0 & 0 & 0 \\
0 & 0 & 0 & 0 & 0 & 0 & 0 & 0 \\
\frac{3}{2} & 0 & 0 & \frac{3}{2} & 3-3a^2 & -3ab & -3ab & 3a^2
\end{bmatrix}$$
Then, for $\rho$ density concentrated on vertex 2,
$$\Big[D\Big(I-G+G_dE\Big)\Big]_{12}vec(\rho)=
\begin{bmatrix}
\frac{3a^2-1}{2b^2} & \frac{a}{b} & \frac{a}{b} & \frac{1}{2}\\
 0 & 0 & 0 & 0 \\
 0 & 0 & 0 & 0 \\
\frac{3}{2} & 0 & 0 & \frac{3}{2}\end{bmatrix}\begin{bmatrix} \rho_{11} \\ \rho_{12} \\ \ov{\rho_{12}}\\ \rho_{22}\end{bmatrix}=\begin{bmatrix} \frac{3a^2-1}{2b^2}\rho_{11}+\frac{\rho_{22}}{2}+\frac{2a}{b}Re(\rho_{12})\\0 \\0 \\\frac{3}{2}(\rho_{11}+\rho_{22})\end{bmatrix}$$
from which we conclude $\mathrm{Tr}([D(I-G+G_dE)]_{12}\rho)$ equals the RHS of (\ref{mhtex1}), as expected. 

\eex
\qee

\bex\label{exemplo2}
Consider the following nonunital QMC $T$ on two vertices determined by
\beq\label{ex1}
\Phi_{11}=V_1\cdot V_1^\dagger+V_2\cdot V_2^\dagger,\;\;\;\Phi_{21}=V_3\cdot V_3^\dagger+V_4\cdot V_4^\dagger,\;\;\;\Phi_{12}=L\cdot L^\dagger,\;\;\;\Phi_{22}=R\cdot R^\dagger
\eeq
$$V_1=\sqrt{1-\frac{3p}{4}}I,\;\;\;V_2=\frac{\sqrt{p}}{2}\begin{bmatrix} 0 & 1 \\ 1 & 0 \end{bmatrix},\;\;\;V_3=\frac{\sqrt{p}}{2}\begin{bmatrix} 0 & -i \\ i & 0 \end{bmatrix},\;\;\;V_4=\frac{\sqrt{p}}{2}\begin{bmatrix} 1 & 0 \\ 0 & -1 \end{bmatrix},$$
$$L=\frac{1}{\sqrt{3}}\begin{bmatrix} 1 & 1 \\ 0 & 1\end{bmatrix},\;\;\;R=\frac{1}{\sqrt{3}}\begin{bmatrix} 1 & 0 \\ -1  & 1\end{bmatrix}$$
where we assume $0<p<1$ (Figure \ref{fig2}). The matrix representations are order 4 matrices and the block matrix representation for $T$ is
$$
\hat{T}=\begin{bmatrix} [\Phi_{11}] & [\Phi_{12}] \\ [\Phi_{21}] & [\Phi_{22}]\end{bmatrix}=\begin{bmatrix} 
1-\frac{3p}{4} & 0 & 0 & \frac{p}{4} & \frac{1}{3} & \frac{1}{3} & \frac{1}{3} & \frac{1}{3}\\
0 & 1-\frac{3p}{4} & \frac{p}{4} &  0 & 0 &  \frac{1}{3} &  0 & \frac{1}{3} \\
0 & \frac{p}{4} & 1-\frac{3p}{4} & 0 & 0 & 0 & \frac{1}{3} & \frac{1}{3} \\
\frac{p}{4} & 0 & 0 & 1-\frac{3p}{4} & 0 & 0 & 0 & \frac{1}{3} \\
\frac{p}{4} & 0 & 0 & \frac{p}{4} & \frac{1}{3} & 0 & 0 & 0 \\
0 & -\frac{p}{4} & -\frac{p}{4} & 0 & -\frac{1}{3}  &  \frac{1}{3} & 0 & 0 \\
0 & -\frac{p}{4} & -\frac{p}{4} & 0 & -\frac{1}{3}  & 0 & \frac{1}{3} & 0 \\
\frac{p}{4} & 0 & 0 & \frac{p}{4} & \frac{1}{3} & -\frac{1}{3} & -\frac{1}{3} & \frac{1}{3}
\end{bmatrix}$$

\begin{center}
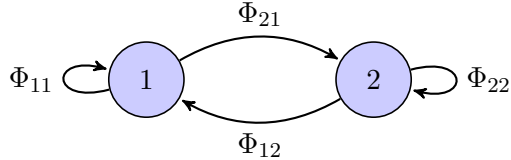
\begin{figure}[ht]
\begin{tikzpicture}
[->,>=stealth',shorten >=1pt,auto,node distance=2.0cm,
                    semithick]
    \node[main node] (1) {$1$};
    \node[main node] (2) [right = 2.0cm and 2.0cm of 1]  {$2$};


    \path[draw,thick]
    (1) edge   [loop left]     node {$\Phi_{11}$} (1)


    (1) edge    [bend left] node[above] {$\Phi_{21}$}    (2)
    (2) edge   [bend left]     node[below] {$\Phi_{12}$} (1)

    (2) edge   [loop right]     node {$\Phi_{22}$ } (2)







    ;
\end{tikzpicture}
\caption{Graph with 2 vertices, with its associated transition maps $\Phi_{ij}$ leading to an ergodic, nonunital QMC, for every $0<p<1$. In general the mean hitting times for reaching a vertex depend on $p$ and the initial density $\rho$. The exact values can be obtained by any of the methods discussed in the previous sections.}\label{fig2}
\end{figure}
\end{center}

Let $\mathbb{P}_i$ denote the projection on (the space generated by) vertex $i$,
\beq
\mathbb{P}_1=\begin{bmatrix} I_4 & 0_4 \\ 0_4 & 0_4\end{bmatrix},\;\;\;\mathbb{P}_2=\begin{bmatrix} 0_4 & 0_4 \\ 0_4 & I_4\end{bmatrix},\;\;\;\mathbb{Q}_i=\mathbb{I}-\mathbb{P}_i,\;\;\;i=1,2\eeq
where $I_4$ and $0_4$ denote the order 4 identity and zero matrices, respectively. Then, for instance, by letting $\rho=\rho_1\otimes|1\rangle\langle 1|$ we have that $\mathrm{Tr}(\mathbb{P}_2\hat{T}\rho)=\mathrm{Tr}(\Phi_{21}\rho_1)$ denotes the probability that the walk will be at vertex $2$ after one step given that it has started at vertex 1 with internal degree $\rho_1$. Regarding its spectrum, $\hat{T}$ has 1 as the unique eigenvalue with unit modulus. By formula (\ref{fixformula}), the unique fixed point is faithful and given by
$$\pi=\pi_1\otimes|1\rangle\langle 1|+\pi_2\otimes|2\rangle\langle 2|=\frac{1}{32+45p}\Bigg(\begin{bmatrix} 13 & 12 \\ 12 & 19\end{bmatrix}\otimes |1\rangle\langle 1|+\begin{bmatrix} 12p & -15p \\ -15p & 33p\end{bmatrix}\otimes|2\rangle\langle 2|\Bigg)$$
Let 
$$\Omega=|\pi\rangle\langle e_{I_2^2}|=\frac{1}{32+45p}\begin{bmatrix} 
13 & 0 & 0 & 13 & 13 & 0 & 0 & 13 \\
12 & 0 & 0 & 12 & 12 & 0 & 0 & 12 \\
12 & 0 & 0 & 12 & 12 & 0 & 0 & 12 \\
19 & 0 & 0 & 19 & 19 & 0 & 0 & 19 \\
12p & 0 & 0 & 12p & 12p & 0 & 0 & 12p \\
-15p & 0 & 0 & -15p & -15p & 0 & 0 & -15p \\
-15p & 0 & 0 & -15p & -15p & 0 & 0 & -15p \\
33p & 0 & 0 & 33p & 33p & 0 & 0 & 33p \end{bmatrix}
$$
The fundamental matrix $Z$ can be obtained explicitly as a funcion of $p$, but we omit its expression here for brevity. In this example every entry $Z_{ij}$ can be written as a rational function of $p$ with numerator and denominator being polynomials of degree at most 2 and 3 respectively. Also,
$$\hat{h}_{11}=p\begin{bmatrix} 
\frac{1}{p}-\frac{1}{2} & -\frac{1}{8} & -\frac{1}{8} & \frac{1}{2}\\
\frac{3}{16} & \frac{1}{p}-\frac{3}{4} & \frac{1}{4} & \frac{3}{16} \\
\frac{3}{16} & \frac{1}{4} & \frac{1}{p}-\frac{3}{4} & \frac{3}{16} \\
\frac{1}{2} & \frac{1}{8} & \frac{1}{8} & \frac{1}{p}-\frac{1}{2}\end{bmatrix},\;\;\;\hat{h}_{12}=\frac{1}{4}\begin{bmatrix} 
2 & 1 & 1 & 2 \\
1 & 1 & -1 & 2 \\
1 & -1 & 1 & 2 \\
2 & -1 & -1 & 2\end{bmatrix}
$$
$$\hat{h}_{21}=\frac{1}{2}\begin{bmatrix} 1 & 0 & 0 & 1 \\
0 & -1 & -1 & 0 \\
0 & -1 & -1 & 0 \\
1 & 0 & 0 & 1\end{bmatrix},\;\;\;\hat{h}_{22}=\frac{1}{6}\begin{bmatrix} 3 & 1 & 1 & 2\\
-2 & 1 & -1 & -2\\
-2 & -1 & 1 & -2\\
3 & -1 & -1 & 4\end{bmatrix}
$$
Then the hitting probabilities are all equal to 1, as expected. Also,
$$\hat{k}_{11}=p\begin{bmatrix}
\frac{1}{p}-\frac{1}{32} & -\frac{1}{8} & -\frac{1}{8}  & \frac{31}{32} \\
\\
\frac{21}{32} & \frac{1}{p}-\frac{9}{16}  & \frac{7}{16} & \frac{21}{32} \\
\\
\frac{21}{32} &  \frac{7}{16} & \frac{1}{p}-\frac{9}{16} & \frac{21}{32} \\
\\
\frac{37}{32} & \frac{1}{2} & \frac{1}{2} & \frac{1}{p}+\frac{5}{32}
\end{bmatrix},\;\;\;\hat{k}_{12}=\frac{1}{4}\begin{bmatrix} 
\frac{9}{2} & 0 & 0 & 3\\

\\
\frac{9}{2} & 0 & -3 & 3 \\
\\

\frac{9}{2} &  -3 & 0 & 3 \\
\\
\frac{15}{2} & -3 & -3 & 3\end{bmatrix}
$$
$$\hat{k}_{21}=\frac{1}{p}\begin{bmatrix} 1 & 0 & 0 & 1 \\ 0 & -1 & -1 & 0 \\ 0 & -1 & -1 & 0 \\1 & 0 & 0 & 1\end{bmatrix},\;\;\;\hat{k}_{22}=\frac{1}{6p}\begin{bmatrix} 
3p+2 & p+2 & p+2 & p+2 \\
-2p & p-2 & -(p+2) & -2(p+2)\\
-2p & -(p+2) & p-2 & -2(p+2)\\
3p+2 & -(p-2) & -(p-2) & \frac{1}{4}(p+1)
\end{bmatrix}
$$
From this we obtain, for $\rho=(\rho_{ij})_{i,j=1,2}$ density matrix,
$$\mathrm{Tr}(\hat{k}_{11}\rho)=1+p\Big[\frac{9}{8}+\frac{3}{4}Re(\rho_{12})\Big],\;\;\; \mathrm{Tr}(\hat{k}_{12}\rho)=3\Big[\rho_{11}+\frac{1}{2}\rho_{22}-\frac{1}{2}Re(\rho_{12})\Big],$$
$$\mathrm{Tr}(\hat{k}_{21}\rho)=\frac{2}{p},\;\;\; \mathrm{Tr}(\hat{k}_{22}\rho)=\frac{1}{3p}[2+3p+2\rho_{22}+4Re(\rho_{12})]$$
A long but routine calculation allows us to check formula (\ref{aldous_eqnova0}) and obtain, as expected, the same results via matrix $Z$. All such calculations are easily performed by a computer but we omit the intermediate steps for brevity since, in this example, the practical calculation with $Z$ seems to be more complicated than by proceeding via the definition of $\hat{k}_{ij}$. Because of this, it is natural to ask what happens with other generalized inverses, so we could proceed via formula (\ref{hunterf2}). As in the previous example, by choosing $|u\rangle=[1\; 0 \; 0\; 0 \;0 \;0 \;0 \;0]^T$ we can define $G=(I-T+|u\rangle\langle e_{I}|)^{-1}$ which, in a similar way as for $Z$, its entries can be written as a rational function of $p$ but this time with numerator and denominator being polynomials of degree at most 1 and 2 respectively. A calculation shows that, in fact, $\hat{G}_{ii}=\hat{k}_{ii}$, $i=1,2$, but that $\hat{G}_{ij}\neq \hat{k}_{ij}$ if $i\neq j$. We have
$$\hat{G}_{21}|\rho\rangle=\frac{1}{p(32+45p)}\begin{bmatrix} 
\frac{1}{2}(9p^2+46p+64) & 6(3p+4) & 6(3p+4) & -\frac{1}{2}(9p^2-50p-64)\\
-2(9p+8) & \frac{3}{4}(15p+8) & -\frac{5}{4}(9p+8) & -2(16+21p)\\
-2(9p+8) & -\frac{5}{4}(9p+8) & \frac{3}{4}(15p+8) & -2(16+21p)\\
-\frac{1}{2}(9p^2-134p-64) & -6(3p+4) & -6(3p+4) & \frac{1}{2}(9p^2+130p+64)
\end{bmatrix}\begin{bmatrix}\rho_{11} \\ \rho_{12} \\ \rho_{21} \\ \rho_{22}\end{bmatrix}$$
$$=\begin{bmatrix}\rho_{11}' & \rho_{12}' & \rho_{21}' & \rho_{22}'\end{bmatrix}^T$$
where
$$\rho_{11}'=\frac{1}{2p(32+45p)}\Bigg[(9p^2+46p+64)\rho_{11}+(50p+64-9p^2)\rho_{22}+2p(36p+48)Re(\rho_{12})\Bigg]$$
$$\rho_{22}'=-\frac{1}{2p(32+45p)}\Bigg[(9p^2-134p-64)\rho_{11}+(-9p^2-130p-64)\rho_{22} +2p(36p+48)Re(\rho_{12})\Bigg]$$
from which we obtain, as expected, that $\mathrm{Tr}(\hat{G}_{21}\rho)=\rho_{11}'+\rho_{22}'=\mathrm{Tr}(\hat{k}_{21}\rho)=2/p$, for every $\rho$ density. A similar procedure allows us to confirm that $\mathrm{Tr}(\hat{G}_{21}\rho)=\mathrm{Tr}(\hat{k}_{12}\rho)$ for every $\rho$ density.

\eex\qee

\bex\label{ex83}

In this set of examples we illustrate the Random Target Lemma, stated in Theorem \ref{oqwmhtfstat}(b). We note that in some special cases such quantity makes sense even if the assumption of such item is not valid, but then we will have a result for a subset of densities only. This notion can be seen as a kind of classical character of the walk, a property which is not possessed by examples in general. We also check the second hitting time formula, as such results are better exemplified via graphs with at least 3 vertices (Figure \ref{fig3}). Here we are able to see how a single pair of matrices can lead to several distinct situations.

\begin{center}
\begin{figure}[ht]
\begin{tikzpicture}
[->,>=stealth',shorten >=1pt,auto,node distance=2.0cm,
                    semithick]
    \node[main node] (1) {$1$};
    \node[] (1a) [below = 2.0cm and 2.0cm of 1,label={}]  {};
    \node[main node] (2) [left =2.0cm and 2.0cm of 1a,label={[]$$}]  {$2$};
    \node[main node] (3) [right = 2.0cm and 4.0cm of 2] {$3$};


    \path[draw,thick]
    (1) edge   [loop above]     node {$B_{11}$} (1)


    (1) edge     node {$B_{21}$}    (2)
    (2) edge   [bend left]     node{$B_{12}$} (1)

    (2) edge   [loop left]     node {$B_{22}$ } (2)
    (2) edge        node {$B_{32}$} (3)
    (3) edge   [bend left]     node[below] {$B_{23}$} (2)

    (3) edge   [loop right]     node {$B_{33}$} (3)

    (3) edge      node {$B_{13}$} (1)
    (1) edge   [ bend left]     node {$B_{31}$} (3)



    ;
\end{tikzpicture}
\caption{Complete graph with 3 vertices and associated transitions $B_{ij}$ from vertex $j$ to vertex $i$, where $i,j=1,2,3$. Examples \ref{ex83}a,b,c are all ergodic but only a and b are unital.
}\label{fig3}
\end{figure}
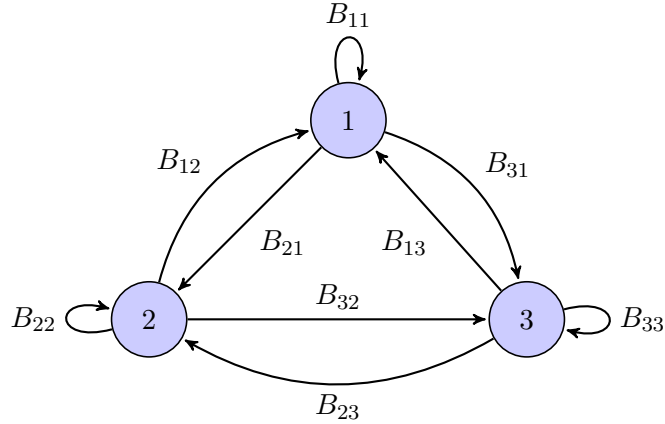
\end{center}

a)  Let
$$B_{11}=B_{22}=B_{33}=\frac{1}{2}\begin{bmatrix} 1 & 1 \\ 0 & 0 \end{bmatrix},\;\;\;B_{13}=B_{32}=B_{21}=\frac{1}{2}\begin{bmatrix} 0 & 0 \\ 1 & -1\end{bmatrix}\;\;\;B_{12}=B_{23}=B_{31}=\frac{1}{\sqrt{2}}I$$
and consider the ergodic unital OQW on 3 vertices with transition effects given by the $B_{ij}$, that is, with block representation matrix and limit operator
$$\hat{T}:=\begin{bmatrix} [B_{11}] & [B_{12}] & [B_{13}] \\ [B_{21}] & [B_{22}] & [B_{13}] \\ [B_{31}] & [B_{32}] & [B_{33}]\end{bmatrix},\;\;\;\hat{\Omega}=|\pi\rangle\langle e_I|=\begin{bmatrix} \Omega_{11} & \Omega_{11} & \Omega_{11} \\ \Omega_{11} & \Omega_{11} & \Omega_{11} \\ \Omega_{11} & \Omega_{11} & \Omega_{11} \end{bmatrix},\;\;\;\Omega_{11}=\frac{1}{6}\begin{bmatrix} 1 & 0 & 0 & 1 \\ 0 & 0 & 0 & 0 \\ 0 & 0 & 0 & 0 \\ 1 & 0 & 0 & 1 \end{bmatrix}$$
Also,
$$\hat{h}_{ii}=\begin{bmatrix} \frac{9}{16} & \frac{9}{32} & \frac{9}{32} & \frac{7}{16}\\
0 & \frac{1}{8} & 0 & 0 \\
0 & 0 & \frac{1}{8} & 0\\
\frac{7}{16} & -\frac{9}{32} & -\frac{9}{32} & \frac{9}{16}\end{bmatrix},\;\;\;\hat{k}_{ii}=\begin{bmatrix} \frac{25}{16} & \frac{13}{32} & \frac{13}{32} & \frac{19}{16}\\
0 & \frac{3}{8} & 0 & 0 \\
0 & 0 & \frac{3}{8} & 0\\
\frac{23}{16} & -\frac{21}{32} & -\frac{21}{32} & \frac{29}{16}\end{bmatrix},\;\;\;i=1,2,3$$

$$\hat{h}_{12}=\hat{h}_{23}=\hat{h}_{31}=\begin{bmatrix} \frac{3}{4} & \frac{1}{8} & \frac{1}{8} & \frac{1}{4}\\
0 & \frac{1}{2} & 0 & 0 \\
0 & 0 & \frac{1}{2} & 0\\
\frac{1}{4} & -\frac{1}{8} & -\frac{1}{8} & \frac{3}{4}\end{bmatrix},\;\;\;\hat{k}_{12}=\hat{k}_{23}=\hat{k}_{31}=\begin{bmatrix} \frac{19}{14} & \frac{11}{56} & \frac{11}{56} & \frac{6}{7}\\
0 & \frac{1}{2} & 0 & 0 \\
0 & 0 & \frac{1}{2} & 0\\
\frac{13}{14} & -\frac{19}{56} & -\frac{19}{56} & \frac{10}{7}\end{bmatrix}$$

$$\hat{h}_{13}=\hat{h}_{21}=\hat{h}_{32}=\begin{bmatrix} \frac{1}{2} & \frac{3}{16} & \frac{3}{16} & \frac{1}{4}\\
0 & \frac{1}{4} & 0 & 0 \\
0 & 0 & \frac{1}{4} & 0\\
\frac{1}{2} & -\frac{3}{16} & -\frac{3}{16} & \frac{3}{4}\end{bmatrix},\;\;\;\hat{k}_{13}=\hat{k}_{21}=\hat{k}_{32}=\begin{bmatrix} \frac{11}{7} & \frac{19}{28} & \frac{19}{28} & \frac{15}{14}\\
0 & \frac{1}{2} & 0 & 0 \\
0 & 0 & \frac{1}{2} & 0\\
\frac{9}{7} & -\frac{1}{28} & -\frac{1}{28} & \frac{25}{14}\end{bmatrix}$$
$$\hat{Z}_{ii}=\begin{bmatrix}    
\frac{15}{14} & \frac{61}{147} & \frac{61}{147} & -\frac{1}{14} \\
0 & \frac{8}{7} & 0 & 0 \\
0 & 0 & \frac{8}{7} & 0\\
-\frac{1}{6} & -\frac{47}{147} & -\frac{47}{147} & \frac{41}{42}
\end{bmatrix},\;\;\;i=1,2,3$$
$$\hat{Z}_{12}=\hat{Z}_{23}=\hat{Z}_{31}=\begin{bmatrix}
\frac{17}{42} & \frac{55}{147} & \frac{55}{147} & -\frac{1}{6} \\
0 & \frac{4}{7} & 0 & 0 \\
0 & 0 & \frac{4}{7} & 0\\
-\frac{11}{42} & -\frac{41}{147} & -\frac{41}{147} & \frac{13}{42}
\end{bmatrix},\;\;\;\hat{Z}_{13}=\hat{Z}_{32}=\hat{Z}_{21}=\begin{bmatrix}
\frac{1}{42} & \frac{31}{147} & \frac{31}{147} & -\frac{11}{42} \\
0 & \frac{2}{7} & 0 & 0 \\
0 & 0 & \frac{2}{7} & 0\\
-\frac{1}{14} & -\frac{59}{147} & -\frac{59}{147} & \frac{3}{14}
\end{bmatrix}$$
A routine calculations gives
$$\mathrm{Tr}(\hat{k}_{12}\rho)=\mathrm{Tr}(\hat{k}_{23}\rho)=\mathrm{Tr}(\hat{k}_{31}\rho)=\frac{16}{7}-\frac{2}{7}Re(\rho_{12})$$
$$\mathrm{Tr}(\hat{k}_{13}\rho)=\mathrm{Tr}(\hat{k}_{32}\rho)=\mathrm{Tr}(\hat{k}_{21}\rho)=\frac{20}{7}+\frac{9}{7}Re(\rho_{12})$$
Now we note that assumption of Theorem \ref{oqwmhtfstat}(b) is not valid, since
$$\mathrm{Tr}(\hat{k}_{ii}\rho)=3(\gamma_{11}+\gamma_{22})-\frac{1}{4}(\gamma_{12}+\gamma_{21})$$
However, we still have that (\ref{aldous_eqnova}) is valid for any diagonal density matrix as a simple inspection shows, so with respect to such densities we have the associated target time
$$t_{\odot}(\rho_{diag})=\sum_{i=1}^3 \mathrm{Tr}[({D}^{-1}{K})_{ij}\rho_{diag}]=\Big[\sum_{i=1}^3 \mathrm{Tr}(Z_{ii}\rho_{diag})\Big]-1=\frac{12}{7},\;\;\;j=1,2,3
$$
As discussed before this is the expected length of the walk, given a randomly chosen target vertex $i$, beginning at some fixed vertex with any diagonal density $\rho_{diag}$. In order to illustrate such vertex independence in a situation which is not as symmetric as this one, see example c) below. As for the second hitting time formula, we inspect both sides of such equation to illustrate their equality. Noting that in this example all $\pi_i$ are equal, we obtain
$$(\hat{k}_{11}+\hat{k}_{12}+\hat{k}_{13})\pi_i=\begin{bmatrix}
\frac{503}{112} & \frac{41}{32} & \frac{41}{32} & \frac{349}{112}\\
0 & \frac{11}{8} & 0 & 0 \\
0 & 0 & \frac{11}{8} & 0\\
\frac{409}{112} & -\frac{33}{32} & -\frac{33}{32} & \frac{563}{112}
\end{bmatrix}\begin{bmatrix} 1/6 \\ 0 \\ 0 \\ 1/6\end{bmatrix}=\begin{bmatrix} \frac{71}{56} \\ 0 \\0 \\ \frac{81}{56}\end{bmatrix}$$
$$(DZ)_{11}(\hat{h}_{11}+\hat{h}_{12}+\hat{h}_{13})\pi_i=\begin{bmatrix}
\frac{439}{112} & \frac{43}{48} & \frac{43}{48} & \frac{397}{112}\\
0 & \frac{3}{8} & 0 & 0 \\
0 & 0 & \frac{3}{8} & 0\\
\frac{473}{112} & -\frac{43}{48} & -\frac{43}{48} & \frac{515}{112}
\end{bmatrix}\begin{bmatrix} 1/6 \\ 0 \\ 0 \\ 1/6\end{bmatrix}=\begin{bmatrix} \frac{209}{168} \\ 0 \\ 0 \\ \frac{247}{168}\end{bmatrix}$$
Therefore, by (\ref{mhtf2eq}),
$$\mathrm{Tr}(K_{j\pi})=\sum_i \mathrm{Tr}(\hat{k}_{ji}\pi_i)=\mathrm{Tr}((D Z)_{jj}F_{1\pi})=\sum_i \mathrm{Tr}((DZ)_{jj}\hat{h}_{ji}\pi_i)=\frac{19}{7},\;\;\;j=1,2,3$$
which is the mean time of first visit to $j$, given that it has started at a vertex randomly chosen according to $\pi$.

\medskip

b) Consider a slight variation of the above, namely
$$B_{11}=B_{22}=B_{33}=\frac{1}{\sqrt{2}}I,\;\;\;B_{13}=B_{32}=B_{21}=\frac{1}{2}\begin{bmatrix} 0 & 0 \\ 1 & -1\end{bmatrix}\;\;\;B_{12}=B_{23}=B_{31}=\frac{1}{2}\begin{bmatrix} 1 & 1 \\ 0 & 0 \end{bmatrix}$$
and consider the OQW on 3 vertices with transition effects given by the $B_{ij}$. Once again this is an ergodic unital channel, but now the assumption of Theorem \ref{oqwmhtfstat}(b) holds: we have that
$$\hat{k}_{ii}|\gamma\rangle=\begin{bmatrix} \frac{7}{4} & -\frac{7}{36} & -\frac{7}{36} & \frac{5}{4}\\
0 & \frac{1}{2} & 0 & 0 \\
0 & 0 & \frac{1}{2} & 0\\
\frac{5}{4} & \frac{7}{36} & \frac{7}{36} & \frac{7}{4}\end{bmatrix}\begin{bmatrix} \gamma_{11} \\ \gamma_{12} \\ \gamma_{21} \\ \gamma_{22}\end{bmatrix}=\begin{bmatrix} \frac{7}{4}\gamma_{11}+\frac{5}{4}\gamma_{22}-\frac{7}{36}(\gamma_{12}+\gamma_{21}) \\ \frac{1}{2}\gamma_{12} \\ \frac{1}{2}\gamma_{21} \\ \frac{5}{4}\gamma_{11}+\frac{7}{4}\gamma_{22}+\frac{7}{36}(\gamma_{12}+\gamma_{21}) \end{bmatrix},\;\;\;i=1,2,3$$
from which we obtain $\mathrm{Tr}(\hat{k}_{ii}\gamma)=3\mathrm{Tr}(\gamma)$ for every $\gamma\in M_2(\mathbb{C})$, $i=1,2,3$. Then, for every $\rho$ density, by choosing $j=1$,
$$t_{\odot}(\rho)=\sum_{i=1}^3 \mathrm{Tr}[({D}^{-1}{K})_{i1}\rho]=0+\frac{4}{3}[1+Re(\rho_{12})]+\frac{4}{3}[1-Re(\rho_{12})]=\frac{8}{3}$$
and the same result is obtained by choosing $j=2,3$, as expected. This can also be verified via the block diagonal of $Z$, as in a). As for the second hitting time formula,
$$\mathrm{Tr}(K_{1\pi})=\mathrm{Tr}((D Z)_{11}F_{1\pi})=\mathrm{Tr}((DZ)_{11}\hat{h}_{11}\pi_1)+\mathrm{Tr}((DZ)_{11}\hat{h}_{12}\pi_2)+\mathrm{Tr}((DZ)_{11}\hat{h}_{13}\pi_3)=\frac{11}{3}$$
and similarly $\mathrm{Tr}(K_{2\pi})=\mathrm{Tr}(K_{3\pi})=11/3$.

\medskip

c) Still considering the above matrices, we can make a modification in order to obtain a less symmetric, nonunital OQW. Let
$$B_{11}=B_{22}=\frac{1}{\sqrt{5}}I,\;\;\;B_{33}=\sqrt{\frac{4}{5}}I,\;\;\;B_{21}=B_{32}=\sqrt{\frac{2}{5}}\begin{bmatrix} 0 & 0 \\ 1 & -1\end{bmatrix},$$
$$B_{13}=B_{23}=\frac{1}{\sqrt{10}}I,\;\;\;B_{31}=B_{12}=\sqrt{\frac{2}{5}}\begin{bmatrix} 1 & 1 \\ 0 & 0 \end{bmatrix}$$
The unique stationary density for the associated OQW is
$$\pi=\pi_1\otimes|1\rangle\langle 1|+\pi_2\otimes|2\rangle\langle 2|+\pi_3\otimes|3\rangle\langle 3|=\begin{bmatrix} \frac{1}{8} & 0 \\ 0 & \frac{1}{24}\end{bmatrix}\otimes|1\rangle\langle 1|+\begin{bmatrix} \frac{1}{24} & 0 \\ 0 & \frac{1}{8}\end{bmatrix}\otimes|2\rangle\langle 2|+\begin{bmatrix} \frac{1}{3} & 0 \\ 0 & \frac{1}{3}\end{bmatrix}\otimes|3\rangle\langle 3|$$
In a similar way as in item a), (\ref{aldous_eqnova}) is valid for any diagonal density matrix. Then we calculate its target time:
$$t_{\odot}(\rho_{diag})=\sum_{i=1}^3 \mathrm{Tr}[({D}^{-1}{K})_{ij}\rho_{diag}]=\Big[\sum_{i=1}^3 \mathrm{Tr}(Z_{ii}\rho_{diag})\Big]-1=\frac{5}{2},\;\;\;j=1,2,3$$
As for the second hitting time formula, we inspect both sides,
$$\hat{k}_{11}\pi_1+\hat{k}_{12}\pi_2+\hat{k}_{13}\pi_3=\begin{bmatrix} \frac{265}{72} \\ 0 \\ 0 \\ \frac{227}{72}\end{bmatrix},\;\;\;(DZ)_{11}(\hat{h}_{11}\pi_1+\hat{h}_{12}\pi_2+\hat{h}_{13}\pi_3)=\begin{bmatrix}\frac{21}{8} \\ 0 \\ 0 \\ \frac{101}{24}\end{bmatrix}$$
Hence, both traces are equal to $\mathrm{Tr}(K_{1\pi})=41/6$. Similarly, $\mathrm{Tr}(K_{2\pi})=\mathrm{Tr}(K_{1\pi})$ and $\mathrm{Tr}(K_{3\pi})=11/6$.

\eex
\qee

\section{Outlook}\label{sec6}

In this work we have discussed generalized inverses associated with QMCs acting on a finite dimensional space and we focused on describing some of the the basic constructions needed to study a notion of mean hitting time. 
Inspired by a collection of results which are valid for stochastic matrices we realize that several of them admit a version which is valid for the class of CP maps studied here, thus leading to applications in a quantum setting. We have also concluded that one is able to develop concrete calculations with relative ease, thus leading to a basic theory with a considerable collection of examples.

\medskip

The following are natural topics of discussion:

\begin{enumerate}
\item {\it Hitting times for states.} What can be said about the mean hitting time to a {\it pure state}? On first sight, it is not clear whether a hitting time formula such as (\ref{clmhtf}) is available in this case. It is instructive to attempt to modify the proof for vertices so that one may recognize the convenience of the block matrix representation when it comes to monitoring the visit to vertices, but much remains to be done in this direction. 
\item {\it Infinite graphs.} Are there analogous results for quantum walks on {\it infinite} graphs? This leads to the problem of studying generalized inverses of operators. This endeavor would most likely have connections with the theory of FR-functions and related generating functions such as the ones seen in \cite{gv}.
\item {\it Unitary evolutions.} Are generalized inverses useful for the study of mean hitting times of iterative {\it unitary} evolutions?  In this situation, the coherences which modify the density matrix contents at every step must be examined in a way which is quite distinct from the analysis of individual paths. If such inverses produce relevant information, do they lead to useful hitting time formulae? A fundamental issue seems evident: if we adopt a linear algebraic approach, what kind of matrix representation one has to employ in the unitary case? 
\end{enumerate}

\medskip

{Acknowledgments.} The author would like to thank the anonymous referees for their careful reading and for several suggestions which led to a marked improvement of the manuscript. We acknowledge financial support from a CAPES/PROAP grant to PPGMat/UFRGS (2018).

\section{Appendix: proofs}

Throughout the proofs we recall the notation $|e_I\rangle=|e_{I_k^n}\rangle$ as $n,k$ are clear from context.

\medskip

{\it Proof of Proposition \ref{hunter33}.} The proof is inspired by \cite{hunter}. We recall that if $M_{ij}$ denotes the $(i,j)$-th minor of a matrix $A$, we define the adjugate matrix as $adj(A):=((-1)^{i+j}M_{ji})_{1\leq i,j\leq n}$, and it holds that
$$A\;adj(A)=adj(A)A=det(A)I$$
By Sylvester's determinant theorem, we have $det(I_m+AB)=det(I_n+BA)$, from which we obtain as a  consequence that
$$det(X+|c\rangle\langle r|)=det(X)+\langle r|\;adj(X)|c\rangle$$
for any vectors $|c\rangle$, $|r\rangle$. Let $\pi$ be the (unique) fixed point of $T $, so $(T -I)\pi =0$ and $T -I$ is singular. Then
$$det(I-T +|t\rangle\langle u|)=det(I-T )+\langle u|adj(I-T )|t\rangle=\langle u|adj(I-T )|t\rangle$$
$$adj(I-T )(I-T )=(I-T )adj(I-T )=det(I-T )I=0$$
If we write $A:=adj(I-T )$ then
$$(I-T )A=0\;\Longrightarrow\; TA=A$$
$$A(I-T )=0\;\Longrightarrow\; AT =A$$
As $T $ has a unique fixed point, $A$ must be of rank 1, so that we can write $A=|v\rangle\langle w|$, and $|v\rangle$ is a multiple of the fixed point of $T $. On the other hand, since $[T^\dagger]=[T]^\dagger$, we have that $AT =A$ implies $T ^\dagger A^\dagger =A^\dagger $ and $A^\dagger =|w\rangle\langle v|$, so $|w\rangle$ is a multiple of the unique fixed point of $T ^\dagger $, which is the identity. For instance, in the case of a QMC with 2 vertices, this corresponds to write
$$|w\rangle=|e_I\rangle:=\begin{bmatrix} 1 & 0 & 0 & 1 & 1 & 0 & 0 & 1\end{bmatrix}^T$$
and analogously for larger graphs. Therefore
$$adj(I-T )=A=|v\rangle\langle w|=k|\pi \rangle\langle e_I|$$

\medskip

Let $\mu_1,\dots, \mu_m$ be the eigenvalues of $A$, $\mu_1=0$. Then $\mathrm{Tr}(adj(A))=\prod_{j=2}^m\mu_j$. As $T $ is irreducible, its eigenvalues $\lambda_1,\dots,\lambda_m$ are such that $\lambda_1=1$ is the only eigenvalue equal to 1, so $\mathrm{Tr}(adj(I-T ))=\prod_{j=2}^m(1-\lambda_j)\neq 0$. But $\mathrm{Tr}(adj(I-T ))=k\mathrm{Tr}(|\pi \rangle\langle e_I|)$, so $k\neq 0$ since $\langle e_I|\pi \rangle$ is always nonzero and thus
$$det(I-T +|t\rangle\langle u|)=k\langle u|\pi \rangle\langle e_I|t\rangle$$
which proves nonsingularity if $\langle u|\pi \rangle\neq 0$ and $\langle e_I|t\rangle\neq 0$.  Now note that
$$(I-T +|t\rangle\langle u|)^{-1}(I-T +|t\rangle\langle u|)=I$$
which implies
\beq\label{eq312} (I-T +|t\rangle\langle u|)^{-1}(I-T )=I-(I-T +|t\rangle\langle u|)^{-1}|t\rangle\langle u|\eeq
But $(I-T )|\pi\rangle=0$, so
$$(I-T +|t\rangle\langle u|)|\pi\rangle=(I-T )|\pi\rangle+|t\rangle\langle u|\pi\rangle=|t\rangle\langle u|\pi\rangle$$
and thus
\beq\label{eq313} \frac{1}{\langle u|\pi\rangle}|\pi\rangle=(I-T +|t\rangle\langle u|)^{-1}|t\rangle\eeq
Replace (\ref{eq313}) into (\ref{eq312}) to obtain
\beq\label{eq314}
 (I-T +|t\rangle\langle u|)^{-1}(I-T )=I-\frac{1}{\langle u|\pi\rangle}|\pi\rangle\langle u|\eeq
Hence
$$ (I-T )(I-T +|t\rangle\langle u|)^{-1}(I-T )=(I-T )-\frac{1}{\langle u|\pi\rangle}(I-T )|\pi\rangle\langle u|=(I-T )$$
showing that $(I-T +|t\rangle\langle u|)^{-1}$ is a $g$-inverse of $I-T $.

\qed

{\it Proof of Corollary \ref{giexps}.} In a similar way as in Proposition \ref{hunter33} we may write
$$(I-T +|t\rangle\langle u|)(I-T +|t\rangle\langle u|)^{-1}=I$$
\beq\label{aeq0}(I-T)(I-T +|t\rangle\langle u|)^{-1}=I-|t\rangle\langle u|(I-T +|t\rangle\langle u|)^{-1}\eeq
Moreover,
$$\langle e_I|(I-T+|t\rangle\langle u|)=\langle e_I|(I-T)+\langle e_I|t\rangle\langle u|=\langle e_I|t\rangle\langle u|$$
which implies
$$
\frac{1}{\langle e_I|t\rangle}\langle e_I|=\langle u|(I-T+|t\rangle\langle u|)^{-1}$$
Insert this into equation (\ref{aeq0}) so
\beq\label{eq318}
(I-T)(I-T +|t\rangle\langle u|)^{-1}=I-\frac{1}{\langle e_I|t\rangle}|t\rangle\langle e_I|\eeq

\medskip

Now set $A=I-T$ and $A^-=(I-T+|t\rangle\langle u|)^{-1}$ in characterization (\ref{eq38}) to obtain
$$A^-+H-A^-AHAA^-$$
$$=(I-T+|t\rangle\langle u|)^{-1}+H-\Bigg[(I-T+|t\rangle\langle u|)^{-1}(I-T)H(I-T)(I-T+|t\rangle\langle u|)^{-1}\Bigg]$$
$$=(I-T+|t\rangle\langle u|)^{-1}+H-\Bigg[\Bigg(I-\frac{1}{\langle u|\pi\rangle}|\pi\rangle\langle u|\Bigg)H  \Bigg(I-\frac{1}{\langle e_I|t\rangle}|t\rangle\langle e_I|  \Bigg)\Bigg]$$
$$=(I-T+|t\rangle\langle u|)^{-1}+H-\Bigg[\Bigg(H-\frac{1}{\langle u|\pi\rangle}|\pi\rangle\langle u|H\Bigg)  \Bigg(I-\frac{1}{\langle e_I|t\rangle}|t\rangle\langle e_I|  \Bigg)\Bigg]$$
$$=(I-T+|t\rangle\langle u|)^{-1}+H-\Bigg[  H- \frac{1}{\langle e_I|t\rangle}H|t\rangle\langle e_I|-\frac{1}{\langle u|\pi\rangle}|\pi\rangle\langle u|H+\frac{1}{\langle u|\pi\rangle\langle e_I|t\rangle}|\pi\rangle\langle u|H|t\rangle\langle e_I|\Bigg]$$
$$=(I-T+|t\rangle\langle u|)^{-1}+\frac{1}{\langle e_I|t\rangle}H|t\rangle\langle e_I|+\frac{1}{\langle u|\pi\rangle}|\pi\rangle\langle u|H-\frac{1}{\langle u|\pi\rangle\langle e_I|t\rangle}|\pi\rangle\langle u|H|t\rangle\langle e_I|$$
where in the second equality we have used (\ref{eq314}) and (\ref{eq318}). Similarly, from characterization (\ref{eq39}) we obtain
$$A^-+(I-A^-A)F+G(I-AA^-)$$
$$=(I-T+|t\rangle\langle u|)^{-1}+\Bigg(I-(I-\frac{1}{\langle u|\pi\rangle}|\pi\rangle\langle u|)\Bigg)F+G\Bigg(I-(I-\frac{1}{\langle e_I|t\rangle}|t\rangle\langle e_I|)\Bigg)$$
$$=(I-T+|t\rangle\langle u|)^{-1}+\frac{1}{\langle u|\pi\rangle}|\pi\rangle\langle u|F+G\frac{1}{\langle e_I|t\rangle}|t\rangle\langle e_I|$$

\medskip

Now we prove c): if
$$G_0=(I-T+|t\rangle\langle u|)^{-1}+\frac{|\pi\rangle\langle u|}{\langle u|\pi\rangle}F+G\frac{|t\rangle\langle e_I|}{\langle e_I|t\rangle}$$
set $F=|\pi\rangle\langle f|$ and $G=|g\rangle\langle e_I|$. Conversely, take
$$\langle f|=\frac{\langle u|F}{\langle u|\pi\rangle},\;\;\;|g\rangle=\frac{G|t\rangle}{\langle e_I|t\rangle}$$
Then
$$G_0=(I-T+|t\rangle\langle u|)^{-1}+H\frac{|t\rangle\langle e_I|}{\langle e_I|t\rangle}+\frac{|\pi\rangle\langle u|}{\langle u|\pi\rangle}H-\frac{|\pi\rangle\langle u|H|t\rangle\langle e_I|}{\langle u|\pi\rangle\langle e_I|t\rangle}$$
set $H=|\pi\rangle\langle f|+|g\rangle\langle e_I|$ and the conclusion follows by a simple inspection. Conversely, set, for instance,
$$\langle f|=\frac{\langle u|}{\langle u|\pi\rangle}H,\;\;\;|g\rangle=\Bigg[I-\frac{|\pi\rangle\langle u|}{\langle u|\pi\rangle}\Bigg]\frac{H|t\rangle}{\langle e_I|t\rangle}$$

\qed

{\it Proof of Proposition \ref{teo41}.} a) Equation $(I-T)|\pi\rangle=0$ has the form $AX=C$, where $X=|\pi\rangle$, $A=I-T$, $C=|0\rangle$. The consistency condition $AA^-C=C$ is clearly satisfied so the general solution is given by
$$|\pi\rangle=(I-(I-T)^-(I-T))|z\rangle=A|z\rangle$$
noting that $z$ must satisfy $\langle e_I|\pi\rangle=1=\langle e_I|Az\rangle$. Let $|v\rangle$ be any vector such that $\langle e_I|Av\rangle\neq 0$ and set $|z\rangle=|v\rangle/\langle e_I|Av\rangle$. Therefore 
$|\pi\rangle$ has the stated form. b)  From map $A$ chosen in Proposition \ref{teo41} we have
$$A=I-\Big[(I-T+|t\rangle\langle u|)^{-1}+|f\rangle\langle e_I|\Big](I-T)=I-(I-T+|t\rangle\langle u|)^{-1}(I-T)-|f\rangle\langle e_I|(I-T)$$
$$=I-I+(I-T +|t\rangle\langle u|)^{-1}|t\rangle\langle u|=(I-T +|t\rangle\langle u|)^{-1}|t\rangle\langle u|$$
the third equality due to (\ref{eq312}) and since $\langle e_I|$ is a fixed point of $T^\dagger $. Substituting such expression in (\ref{fixeq41}) gives
$$|\pi\rangle=\frac{(I-T +|t\rangle\langle u|)^{-1}|t\rangle\langle u|v\rangle}{\langle e_I |(I-T +|t\rangle\langle u|)^{-1}|t\rangle\langle u|v\rangle}=\frac{(I-T +|t\rangle\langle u|)^{-1}|t\rangle}{\langle e_I |(I-T +|t\rangle\langle u|)^{-1}|t\rangle}$$
noting that we can always choose $v$ such that $\langle u|v\rangle\neq 0$ and that the expression in the denominator is nonzero due to (\ref{eq313}).

\qed

{\it Proof of Theorem \ref{bigtheorem}.} a) We begin with the definition $L:=K-(K-D)T$, so that we can write
$$K(I-T)=L-DT$$
By Proposition \ref{corsol1}, we have that
\beq\label{stareq1} K=(L-DT)G+V(I-(I-T)G)\eeq
if the consistency condition holds:
$$(L-DT)G(I-T)=L-DT\;\Longleftrightarrow\; (L-DT)[I-G(I-T)]=0$$
This is in fact true, since
$$(L-DT)[I-G(I-T)]=(L-DT)[I-[(I-T+|t\rangle\langle u|)^{-1}+|\pi\rangle\langle f|+|g\rangle\langle e_I|](I-T)]$$
$$=(L-DT)[I-(I-T+|t\rangle\langle u|)^{-1}(I-T)-|\pi\rangle\langle f|(I-T)-|g\rangle\langle e_I|(I-T)]]$$
$$\stackrel{(\ref{eq314})}{=}(L-DT)\Big[I-I+\frac{|\pi\rangle\langle u|}{\langle u|\pi\rangle}-|\pi\rangle\langle f|(I-T)\Big]=(L-DT)\Big[\frac{|\pi\rangle\langle u|}{\langle u|\pi\rangle}-|\pi\rangle\langle f|(I-T)\Big]$$
$$=(L-DT)|\pi\rangle\Big[\frac{\langle u|}{\langle u|\pi\rangle}-\langle f|(I-T)\Big]=(L|\pi\rangle-D|\pi\rangle)\Big[\frac{\langle u|}{\langle u|\pi\rangle}-\langle f|(I-T)\Big]=0$$
noting that $L|\pi\rangle=D|\pi\rangle$. To see why, just note that as $|\pi\rangle$ is fixed for $T$,
$$L|\pi\rangle=K|\pi\rangle-KT|\pi\rangle+DT|\pi\rangle=K|\pi\rangle-K|\pi\rangle+D|\pi\rangle=D|\pi\rangle$$
Then,
$$V(I-(I-T)G)=V\Big(I-(I-T)(I-T+|t\rangle\langle u|)^{-1}+|\pi\rangle\langle f|+|g\rangle\langle e_I|)\Big)$$
$$=V\Big(I-(I-T)(I-T+|r\rangle\langle u|)^{-1}-(I-T)|\pi\rangle\langle f|-(I-T)|g\rangle\langle e_I|\Big)$$
$$\stackrel{(\ref{eq318})}{=}V\Big(I-I+\frac{|t\rangle\langle e_I|}{\langle e_I|t\rangle}-(I-T)|g\rangle\langle e_I|\Big)=V\Big(\underbrace{\frac{|t\rangle}{\langle e_I|t\rangle}-(I-T)|g\rangle}_{|h\rangle}\Big)\langle e_I|=\underbrace{V|h\rangle}_{|b\rangle}\langle e_I|=|b\rangle\langle e_I|$$
In particular,
\beq\label{lembrete}I-G+TG=|h\rangle\langle e_I|\eeq
and (\ref{stareq1}) can be written as
\beq\label{stareqA} K=(L-DT)G+|b\rangle\langle e_I|\eeq

\medskip

Let us examine such equation. For simplicity we illustrate the case of $2$ vertices, so
$$LG=\begin{bmatrix} L_{11}G_{11}+L_{12}G_{21} & L_{11}G_{12}+L_{12}G_{22} \\ L_{21}G_{11}+L_{22}G_{21} & L_{21}G_{12}+L_{22}G_{22}\end{bmatrix}$$
Then
$$\rho:=\begin{bmatrix} \rho_1 \\ 0\end{bmatrix}\;\Longrightarrow\;LG\rho=\begin{bmatrix} (L_{11}G_{11}+L_{12}G_{21})\rho_1 \\ (L_{21}G_{11}+L_{22}G_{21})\rho_1 \end{bmatrix},\;\;\;G\rho=\begin{bmatrix} G_{11}\rho_1 \\ G_{21}\rho_1\end{bmatrix}$$
Let $\mathrm{Tr}^{(i)}(\gamma)$ denote the trace of the positive map on the $i$-th vertex of an OQW density $\gamma$, $i=1,2$. Then by the positivity of the maps $G_{ij}$ and (\ref{lleq1}), we have
$$\mathrm{Tr}^{(1)}(LG\rho)=\mathrm{Tr}(L_{11}G_{11}\rho_1)+\mathrm{Tr}(L_{12}G_{21}\rho_1)=\mathrm{Tr}(G_{11}\rho_1)+\mathrm{Tr}(G_{21}\rho_1)=\mathrm{Tr}(G\rho)$$
On the other hand, if we write
$$|\pi\rangle=\begin{bmatrix} |\pi_a\rangle \\ |\pi_b\rangle\end{bmatrix},\;\;\;|\pi_a\rangle:=\begin{bmatrix}\pi_1 \\ \pi_2 \\ \pi_3 \\ \pi_4\end{bmatrix},\;\;\;|\pi_b\rangle:=\begin{bmatrix} \pi_5 \\ \pi_6 \\ \pi_7 \\ \pi_8\end{bmatrix}$$
then
$$D|\pi\rangle\langle e_{I_2}|=\begin{bmatrix} \hat{k}_{11} & 0 \\ 0 & \hat{k}_{22}\end{bmatrix}\begin{bmatrix} |\pi_a\rangle\langle e_I| & |\pi_a\rangle\langle e_I| \\ |\pi_b\rangle\langle e_I| & |\pi_b\rangle\langle e_I|\end{bmatrix}=\begin{bmatrix} \hat{k}_{11}|\pi_a\rangle\langle e_I| & \hat{k}_{11}|\pi_a\rangle\langle e_I| \\ \hat{k}_{22}|\pi_b\rangle\langle e_I| & \hat{k}_{22}|\pi_b\rangle\langle e_I|\end{bmatrix}$$
and so
$$D\Omega G\rho=D|\pi\rangle\langle e_{I_2}|G\rho=\begin{bmatrix} k_{11}|\pi_a\rangle\langle e_I|(G_{11}+G_{21}) & k_{11}|\pi_a\rangle\langle e_I|(G_{12}+G_{22}) \\ k_{22}|\pi_b\rangle\langle e_I|(G_{11}+G_{21}) & k_{22}|\pi_b\rangle\langle e_I|(G_{12}+G_{22})\end{bmatrix}\begin{bmatrix} \rho_1 \\ 0\end{bmatrix}$$
$$=\begin{bmatrix} k_{11}|\pi_a\rangle\langle e_I|(G_{11}+G_{21})\rho_1\\ k_{22}|\pi_b\rangle\langle e_I|(G_{11}+G_{21})\rho_1 \end{bmatrix}$$
By Kac's Lemma for OQWs [\cite{bbp}, Thm. 4.5], and noting that $\langle e_I|(G_{11}+G_{21})|\rho_1\rangle=\mathrm{Tr}[(G_{11}+G_{21})\rho_1]$, 
$$\mathrm{Tr}^{(1)}(D\Omega G\rho)=\mathrm{Tr}[(G_{11}+G_{21})\rho_1]=\mathrm{Tr}(G\rho)$$
and we conclude that $\mathrm{Tr}^{(1)}(LG\rho)=\mathrm{Tr}^{(1)}(D\Omega G\rho)$ for every $\rho$. Similarly, by defining $\rho:=[0\;\rho_2]^T$ we conclude $\mathrm{Tr}^{(2)}(LG\rho)=\mathrm{Tr}^{(2)}(D\Omega G\rho)$ and by linearity we conclude that there is $W_\alpha$ such that 
\beq\label{starstar}
LG=D\Omega G+W_\alpha\eeq
and $\mathrm{Tr}(W_\alpha\rho)=0$, for every $\rho$. The case of 3 or more vertices follows in an analogous manner.

\medskip

Now, motivated by a classical reasoning we seek a matrix $B$ such that $\mathrm{Tr}(BE\rho)=\mathrm{Tr}(|b\rangle\langle e_I|\rho)$ for every $\rho$, where $E$ is the block matrix for which each block equals the identity (of order $k^2$, where $k$ is the internal degree of the QMC). For this, we need to find $B_\alpha$ such that
\beq\label{claim1e}
BE+B_\alpha=|b\rangle\langle e_I|,\;\;\;\mathrm{Tr}(B_\alpha\rho)=0,\;\;\;\forall \rho\eeq
We will see that it is enough to find $B$ which is diagonal. For instance, in the case of $n=1$ vertex and degree of freedom $k=2$, by writing $|b\rangle=[b_1\;\cdots\; b_4]^T$, $B=diag(c_1,\dots,c_{4})$ then by writing (\ref{claim1e}) explicitly, $E$ is the order 4 identity matrix, so
$$B_\alpha= |b\rangle\langle e_I|-BE=\begin{bmatrix} b_1-c_1 & 0 & 0 & b_1\\ b_2 & -c_2 & 0 & b_2 \\ b_3 & 0 & -c_3 & b_3 \\ b_4 & 0 & 0 & b_4-c_4\end{bmatrix}\;\Longrightarrow\; Tr(B_\alpha\rho)=(b_1+b_4-c_1)\rho_{11}+(b_1+b_4-c_4)\rho_{22}$$
for $\rho=(\rho_{ij})_{i,j=1,2}$, we conclude that we may set $c_1=c_4=b_1+b_4$, with $b_2,b_3$ arbitrary. In the case of $n=2$ vertices and $k=2$, by writing $B=diag(c_1,\dots,c_{8})$ then we obtain that we may set $c_1=c_4=b_1+b_4$, $c_5=c_8=b_5+b_8$ and the remaining entries are arbitrary. The case of $n$ or $k\geq 3$  follows in a similar way. Therefore we have:

\medskip

{Claim 1:} for every $|b\rangle$ there is $B$ diagonal matrix and $B_\alpha$ such that (\ref{claim1e}) holds.

\medskip

Now write $K_d=D$ and note that
$$(DTG)_d=D(TG)_d$$
From (\ref{stareqA}),
$$K=(L-DT)G+|b\rangle\langle e_I|\;\Longrightarrow D=(LG)_d-(DTG)_d+B=D(\Omega G)_d-D(TG)_d+B+W_\alpha$$
which implies
\beq\label{stareqB}
B=D(I-(\Omega G)_d+(TG)_d)+W_\alpha\eeq
Now replace (\ref{claim1e}) into (\ref{stareqA}):
$$K=(L-DT)G+BE+B_\alpha\stackrel{(\ref{stareqB})}{=}(L-DT)G+\Big(D(I-(\Omega G)_d+(TG)_d)+W_\alpha\Big)E+B_\alpha$$
$$\stackrel{(\ref{starstar})}{=}D\Omega G+W_\alpha-DTG+DE-D(\Omega G)_dE+D(TG)_dE+W_\alpha E+B_\alpha$$
$$=D\Big(\Omega G-(\Omega G)_dE-TG+(TG)_dE+E\Big)+\underbrace{W_\alpha+W_\alpha E+B_\alpha}_{A_\alpha}$$
\beq\label{almost1}
=D\Big(\Omega G-(\Omega G)_dE\Big)+D\Big(E-TG+(TG)_dE\Big)+A_\alpha\eeq

\medskip

{Claim 2:} there is $W_\beta$ such that
$$|h\rangle\langle e_I|=(I-G+TG)_dE+W_\beta$$
and $\mathrm{Tr}(W_\beta\rho)=0$, for every $\rho$. The proof is identical to the one given for Claim 1.

\medskip

By (\ref{lembrete}),
$$I-G+TG=E-G_dE+(TG)_dE+W_\beta$$
which implies that
$$E-TG+(TG)_dE=I-G+G_dE-W_\beta$$
Substitute this into (\ref{almost1}) to conclude that
$$K=D\Big(\Omega G-(\Omega G)_dE+I-G+G_dE\Big)+A_\alpha-DW_\beta$$
By taking the trace of the above applied to a density and noting that the terms outside the parenthesis all lead to null traces, we are done.

\medskip

b) The expression for $G$ above is obtained from the first expression of Corollary \ref{giexps}, namely,
$$G=(I-T+|t\rangle\langle u|)^{-1}+H\frac{|t\rangle\langle e_I|}{\langle e_I|t\rangle}+\frac{|\pi\rangle\langle u|}{\langle u|\pi\rangle}H-\frac{|\pi\rangle\langle u|H|t\rangle\langle e_I|}{\langle u|\pi\rangle\langle e_I|t\rangle}$$
by first exchanging $|t\rangle$ with $|u\rangle$, by setting $|t\rangle=|e_{I_2}\rangle$ and $H=|f\rangle\langle e_{I_2}|$. Now
$$\Omega G=|\pi\rangle\langle e_{I_2}|(I-T+|u\rangle\langle e_{I_2}|)^{-1}+|\pi\rangle\langle e_{I_2}|f\rangle\langle e_{I_2}|=|\pi\rangle\langle e_{I_2}|(I-T+|u\rangle\langle e_{I_2}|)^{-1}+\langle e_{I_2}|f\rangle|\pi\rangle\langle e_{I_2}|$$
But
$$|\pi\rangle\langle e_{I_2}|(I-T+|u\rangle\langle e_{I_2}|)=|\pi\rangle\langle e_{I_2}|(I-T)+\langle e_{I_2}|u\rangle|\pi\rangle\langle e_{I_2}|=\langle e_{I_2}|u\rangle|\pi\rangle\langle e_{I_2}|$$
and so
$$\frac{|\pi\rangle\langle e_{I_2}|}{\langle e_{I_2}|u\rangle}=|\pi\rangle\langle e_{I_2}|(I-T+|u\rangle\langle e_{I_2}|)^{-1}$$
Replacing this expression into the one for $\Omega G$ above gives
$$\Omega G=\frac{|\pi\rangle\langle e_{I_2}|}{\langle e_{I_2}|u\rangle}+\langle e_{I_2}|f\rangle|\pi\rangle\langle e_{I_2}|=\beta\Omega,\;\;\;\beta=\frac{1}{\langle e_{I_2}|u\rangle}+\langle e_{I_2}|f\rangle$$
Finally, note that
$$(\Omega G)_dE=\beta\Omega_dE=\beta\Omega=\Omega G$$
By replacing this in the formula obtained in a), we are done.

\qed

{\it Proof of Theorem \ref{oqwmhtfstat}b).} Note that
$$\mathrm{Tr}( \hat{k}_{ii}\gamma)=c\mathrm{Tr}(\gamma),\;\forall \gamma\;\Longrightarrow\;\mathrm{Tr}( \hat{k}_{ii}( \hat{k}_{ii})^{-1}\rho)=c\mathrm{Tr}(( \hat{k}_{ii})^{-1}\rho)\;\Longrightarrow\; \mathrm{Tr}(( \hat{k}_{ii})^{-1}\rho)=\frac{1}{c}\mathrm{Tr}(\rho),$$
Therefore
$$\mathrm{Tr}((D^{-1}N)_{ij}\rho)=\mathrm{Tr}(( \hat{k}_{ii})^{-1}N_{ij}\rho)=\frac{1}{c}\mathrm{Tr}(N_{ij}\rho)\stackrel{a)}{=}\frac{1}{c}\mathrm{Tr}([(\hat{k}_{ii} \hat{ Z}_{ii}- \hat{k}_{ii} \hat{ Z}_{ij}]\rho)$$
$$=\frac{1}{c}c\mathrm{Tr}([ \hat{ Z}_{ii}- \hat{ Z}_{ij}]\rho)=\mathrm{Tr}([ \hat{ Z}_{ii}- \hat{ Z}_{ij}]\rho)$$
The last statement on the independence of $j$ follows from [\cite{oqwmhtf}, Cor. 2].

\qed

{\it Proof of Theorem \ref{MHTF2}.} This proof is inspired by a classical result \cite{peres}. Let, for any vertices $x,y$,
$$
\hat{P}^n(x,y):=\sum_n\sum_{C\in\mathcal{P}(x\to y;n)}\hat{C}$$
where $\mathcal{P}(x\to y;n)$ denotes the set of products of effect matrices of $T$ associated with all paths of length $n$ moving from vertex $x$ to $y$, and $\hat{C}$ is the conjugation map $\hat{C}\rho:=C\rho C^\dagger$ so by going through all paths $C=B_{i_n i_{n-1}}\cdots B_{i_3 i_2}B_{i_2 i_1}$ allowed by the QMC $T$ we obtain the above operator. The fact that such operator is well-defined follows from the reasoning described, for instance, in \cite{bbp}. Now let
$$\hat{f}_{j}^k:=\sum_i\sum_{C\in\ \mathcal{P}_{fv}(i\to j;k)}\hat{C}\pi_i,\;\;\;F(s):=\sum_{k=0}^\infty \hat{f}_{j}^ks^k$$
where $\mathcal{P}_{fv}(i\to j;k)$ denotes the set of paths from $i$ to $j$ of length $k$ such that the first visit to $j$ occurs at time $k$. Also let
$$u_k:=\hat{P}^k(j,j)-\hat{\Omega}_{jj},\;\;\;U(s):=\sum_{k=0}^\infty u_ks^k,\;\;\;j\in \Omega$$
Then we have
$$
\pi_j=\sum_{k=0}^m \hat{P}^{m-k}(j,j)\hat{f}_{j}^k=\sum_{k=0}^m \Big[\Big(\underbrace{\hat{P}^{m-k}(j,j)-\hat{\Omega}_{jj}\Big)}_{u_{m-k}}+\hat{\Omega}_{jj}\Big]\hat{f}_{j}^k=\sum_{k=0}^m \Big[u_{m-k}+\hat{\Omega}_{jj}\Big]\hat{f}_{j}^k$$
Thus, the constant sequence equal to $\pi_j$ is the convolution of
the sequence of matrices $\{u_k +\hat{\Omega}_{jj}\}_{k=0}^\infty$ with the sequence of vectors $\{\hat{f}_{j}^k\}_{k=0}^\infty$. Then, the generating function of $\{\pi_j\}$
equals the product of the generating functions of  $\{u_k +\hat{\Omega}_{jj}\}_{k=0}^\infty$, which is
$$
\sum_{m=0}^\infty [u_m+\hat{\Omega}_{jj}]s^m=U(s)+\frac{\hat{\Omega}_{jj}}{1-s}$$
and of $\{\hat{f}_{j}^k\}_{k=0}^\infty$ (which is $F(s)$). Therefore,
$$\frac{\pi_j}{1-s}=\sum_{m=0}^\infty \pi_js^m=\Big[U(s)+\frac{\hat{\Omega}_{jj}}{1-s}\Big]F(s) \;\Longrightarrow\; \pi_x=[(1-s)U(s)+\hat{\Omega}_{jj}]F(s)$$
which also holds for $s=1$. Differentiating with respect to $s$ at $s=1$ gives
$$
0=-U(1)F(1)+\hat{\Omega}_{jj}F'(1)$$
and so
$$
\hat{\Omega}_{jj}F'(1)=U(1)F(1)\;\Longrightarrow\; \hat{\Omega}_{jj}F'(1)=\hat{ Z}_{jj}F_{j\pi}$$
Apply $\hat{D}_{jj}$ on the left of both sides and, noting that $\mathrm{Tr}(\hat{D}_{jj}\hat{\Omega}_{jj}\rho)=\mathrm{Tr}(\rho)$ and that $(\hat{D}\hat{ Z})_{jj}=\hat{D}_{jj}\hat{ Z}_{jj}$ we obtain, after taking the trace,
$$\mathrm{Tr}(N_{j\pi})=\mathrm{Tr}[(\hat{D}\hat{ Z})_{jj}F_{j\pi}]$$

\qed


\begin{thebibliography}{99}

\bibitem{accardi2} L. Accardi, D. Koroliuk. Quantum Markov chains: The recurrence problem. Quant. Prob. Rel. Top. VII, 63-73 (1991).

\bibitem{accardi1} L. Accardi, D. Koroliuk. Stopping times for quantum Markov chains. Journ. Th. Prob., Vol. 5, no. 3, pp 521-535, 1992.

\bibitem{aldous} D. Aldous, J. Fill. Reversible Markov Chains and Random Walks on Graphs. \href{http://www.stat.berkeley.edu/~aldous/RWG/book.html}{http://www.stat.berkeley.edu/$\sim$aldous/RWG/book.html}

\bibitem{ambainis} A. Ambainis, E. Bach, A. Nayak, A. Vishwanath, J. Watrous. One-dimensional quantum walks, in: Proceedings of the 33rd Annual ACM Symposium on Theory of Computing, 2001, pp. 60–69.

\bibitem{ambainis2} A. Ambainis. Quantum walks and their algorithm applications. Int. Journ. Quant. Inf., Vol. 01, No. 4, p. 507-518 (2003).


\bibitem{attal} S. Attal, F. Petruccione, C. Sabot, I. Sinayskiy. Open Quantum Random Walks. J. Stat. Phys.  147:832-852 (2012).

\bibitem{attal2} S. Attal, N. Guillotin-Plantard, C. Sabot. Central Limit Theorems for Open Quantum Random Walks and Quantum Measurement Records. Ann. Henri Poincar\'e 16 (2015), 15-43.


\bibitem{bbp} I. Bardet, D. Bernard, Y. Pautrat. Passage times, exit times and Dirichlet problems for open quantum walks. J. Stat. Phys. 167, 173 (2017).


\bibitem{bourg} J. Bourgain, F. A. Gr\"unbaum, L. Vel\'azquez, J. Wilkening. Quantum recurrence of a subspace and operator-valued Schur functions, Comm. Math. Phys., 329 (2014) 1031-1067.


\bibitem{bremaud} P. Br\'emaud. Markov Chains: Gibbs Fields, Monte Carlo Simulation and Queues. Texts in Applied Mathematics 31. Springer, 1999.

\bibitem{breuerp} H. P. Breuer and F. Petruccione. The Theory of Open Quantum Systems. Oxford University Press, Oxford, 2002.


\bibitem{campbell} S. L. Campbell, C. D. Meyer, Jr. Generalized inverses of linear transformations. Pitman, London, 1979.

\bibitem{cp} R. Carbone, Y. Pautrat. Open Quantum Random Walks: Reducibility, Period, Ergodic Properties. Ann. Henri Poincar\'e 17 (2016) 99-135.

\bibitem{cp2} R. Carbone, Y. Pautrat. Homogeneous open quantum random walks on a lattice. J. Stat. Phys. (2015) 160:1125-1153.

\bibitem{cgl} S. L. Carvalho, L. F. Guidi, and C. F. Lardizabal. Site recurrence of open and unitary quantum walks on the line. Quantum Inf. Process. 16:17 (2017).

\bibitem{dhahri} A. Dhahri, F. Mukhamedov, Open quantum random walks, quantum Markov
chains and recurrence. Rev. Math. Phys. DOI: 10.1142/S0129055X1950020X.

\bibitem{ellinas} D. Ellinas, C. Konstandakis. Parametric quantum search algorithm by CP maps:
algebraic, geometric and complexity aspects. J. Phys. A: Math. Theor. 46 (2013) 415303.

\bibitem{snell} C. M. Grinstead, J. L. Snell. Introduction to probability. American Mathematical Society, 1997.


\bibitem{gv} F. A. Gr\"unbaum, L. Vel\'azquez. A generalization of Schur functions: applications to Nevanlinna functions, orthogonal polynomials, random walks and unitary and open quantum walks. Adv. Math. 326 (2018) 352-464.

\bibitem{werner} F. A. Gr\"unbaum, L. Vel\'azquez, A. H. Werner, R. F. Werner. Recurrence for Discrete Time Unitary Evolutions. Comm. Math. Phys. 320, 543-569 (2013).


\bibitem{gudder} S. Gudder. Quantum Markov chains. J. Math. Phys. 49, 072105 (2008).

\bibitem{gudder2} S. Gudder. Analysis of a quantum Markov chain. Ann. Inst. Henri Poincaré. Vol. 52, no. 1, 1990, p. 31-50.

\bibitem{hj2} R. A. Horn, C. R. Johnson. Topics in matrix analysis. Cambridge University Press, 1991.

\bibitem{hunter} J. J. Hunter. Generalized inverses and their application to applied probability problems. Lin. Alg. Appl. 45:157-198 (1982).


\bibitem{kassal} I. Kassal, S. P. Jordan, P. J. Love, M. Mohseni, A. Aspuru-Guzik.  Polynomial-time quantum algorithm for the simulation of chemical dynamics. Proc. Natl. Acad. Sci. 105 (48) 18681-18686 (2008).

\bibitem{kempe}  J. Kempe. Quantum random walks: An introductory overview. Cont. Phys., 44:4, 307-327 (2003).

\bibitem{kempe2} J. Kempe. Discrete Quantum Walks Hit Exponentially Faster. Prob. Th. Rel. Fields 133, 215-235 (2005).

\bibitem{kendon} V. Kendon, B. Tregenna. Decoherence can be useful in quantum walks. Phys.Rev. A67, 042315 (2003).

\bibitem{kitagawa} T. Kitagawa, M. S. Rudner, E. Berg, and E. Demler: Exploring topological phases with quantum walks, Phys. Rev. A, 82, 033429 (2010).

\bibitem{konno} N. Konno. A new type of limit theorems for the one-dimensional quantum random walk. J. Math. Soc. Japan Vol. 57, No. 4 (2005), 1179-1195.


\bibitem{brun0} H. Krovi, T. A. Brun. Quantum walks with infinite hitting times. Phys. Rev. A 74, 042334 (2006)

\bibitem{kummerer} B. K\"ummerer, H. Maassen. A pathwise ergodic theorem for quantum trajectories. J. Phys. A Math. Gen. 37, 11889-11896 (2004).


\bibitem{ls2015} C. F. Lardizabal, R. R. Souza. On a class of quantum channels, open random walks and recurrence. J. Stat. Phys. (2015) 159:772-796.

\bibitem{ls2016} C. F. Lardizabal, R. R. Souza.  Open quantum random walks: ergodicity, hitting times, gambler's ruin and potential theory. J. Stat. Phys. (2016) 164:1122-1156. 

\bibitem{oqwmhtf} C. F. Lardizabal. Open quantum random walks and the mean hitting time formula. Quant. Inf. Comp. Vol. 17, No. 1$\&$2 (2017) 79-105. ArXiv e-prints:1603.06255.


\bibitem{peres} D. A. Levin, Y. Peres, E. L. Wilmer. Markov chains and mixing times. AMS, Providence, RI, 2009.

\bibitem{meyer75} C. D Meyer, Jr. The Role of the Group Generalized Inverse in the Theory of Finite Markov Chains. SIAM Rev., Vol. 17, No. 3 (Jul., 1975), pp. 443-464.

\bibitem{nielsen} M. A. Nielsen, I. L. Chuang. Quantum computation and quantum information. Cambridge University Press, 2000.


\bibitem{gawron} \L. Pawela, P. Gawron, J. A. Miszczak, P. Sadowski. Generalized Open Quantum Walks on
Apollonian Networks. PLoS ONE 10(7):e0130967 (2015).

\bibitem{portugal} R. Portugal. Quantum walks and search algorithms. Springer, 2013.

\bibitem{sadowski} P. Sadowski, L. Pawela. Central limit theorem for reducible and irreducible
open quantum walks. Quantum Inf. Process. (2016) 15:2725-2743.

\bibitem{salvador} S. E. Venegas-Andraca. Quantum walks: a comprehensive review. Quantum Inf. Process. (2012) 11:1015-1106.

\bibitem{sinayskiy0} I. Sinayskiy, F. Petruccione. Efficiency of open quantum walk implementation of dissipative quantum computing algorithms. Quantum Inf. Process. (2012) 11:1301-1309.

\bibitem{sinayskiy4} I. Sinayskiy, F. Petruccione. Microscopic derivation of open quantum Brownian motion: a particular example. Phys. Scr. T165 (2015) 014017.


\bibitem{spsurvey} I. Sinayskiy, F. Petruccione. Open quantum walks. Eur. Phys. J. Spec. Top. DOI: 10.1140/epjst/e2018-800119-5


\bibitem{sutter} D. Sutter. Approximate Quantum Markov Chains. Springer (2018).

\bibitem{verst} F. Verstraete, M.M. Wolf, J.I. Cirac. Quantum computation and quantum-state engineering driven by dissipation. Nat. Phys. 5, 633 (2009).


\bibitem{wang} G. Wang, Y. Wei, S. Qiao. Generalized inverses: theory and computations. Springer 2018.


\end{thebibliography}
\end{document}